 \UseRawInputEncoding
\documentclass[%
reprint,
superscriptaddress,
floatfix,
 amsmath,amssymb,amsfonts,
prb,
]{revtex4-2}

\usepackage{hyperref}
\usepackage{graphicx}
\usepackage{dcolumn}
\usepackage{bm, bbold}
\usepackage{comment, color}
\usepackage{natbib}
\usepackage{tabularx}

\usepackage[utf8]{inputenc}
\usepackage{lmodern}

\usepackage{hyperref}

\def\bea{\begin{eqnarray}}
\def\eea{\end{eqnarray}}
\def\beq{\begin{equation}}
\def\eeq{\end{equation}}
\def\f{\frac}

\def\e{\epsilon}

\def\be{\beta}

\def\h{\theta}
\def\t{\tau}
\def\a{\alpha}

\def\r{\rho}

\def\s{\sigma}

\def\la{\langle}
\def\ra{\rangle}
\def\nn{\nonumber}

\def\d{\delta}

\def\L{\Lambda}

\def\g{\gamma}

\def\rv{ {\mathbf r}}

\def\a{\alpha}
\def\d{\delta}

\def\la{\langle}
\def\ra{\rangle}
\def\e{\epsilon}

\def\g{\gamma}

\begin{document}


\title{Activated solids: Spontaneous deformations, non-affine fluctuations, softening, and failure}


\date{\today}

\author{Parswa Nath}
\affiliation{Institute of Physics, Sachivalaya Marg, Bhubaneswar-751005, Odisha, India}            

\author{Debankur Das}
\affiliation{Georg-August-Universität Göttingen, Wilhelmsplatz 1, 37073 Göttingen, Germany}

\author{Surajit Sengupta}
\email[Deceased]{}
\affiliation{%
 Tata Institute for Fundamental Research Hyderabad, 36/P Gopanapally, Hyderabad 500107,  India.
}%

\author{Debasish Chaudhuri}
\email[Corresponding author:~]
{debc@iopb.res.in}
\affiliation{Institute of Physics, Sachivalaya Marg, Bhubaneswar-751005, Odisha, India}
\affiliation{Homi Bhabha National Institute, Anushakti Nagar, Mumbai 400094, India}
\begin{abstract}
Internal activity can fundamentally reshape the mechanical behavior of solids, yet its role in softening and failure remains incompletely understood. In this study, we investigate spontaneous deformations in activated solids via non-affine fluctuations that quantify local rearrangements relative to global strain. 
Using scaling analysis and numerical simulations, we show that non-affinity in crystalline solids grows quadratically with active speed, increases linearly with persistence time before saturating, and scales inversely with the distance to the melting density. Spatial correlations reveal an activity-dependent growing correlation length, while relaxation dynamics are governed by the active persistence time. With increasing activity, the distributions of local non-affinity broaden, become more skewed, and develop heavy tails, eventually forming a secondary maximum that signals coexisting small and large non-affinities; this heterogeneity precedes defect formation and two-step melting from solid to hexatic and ultimately to fluid. Finally, we demonstrate that spatially patterned activation provides a simple route to locally induce non-affinity and mechanical softening. Our predictions are experimentally testable and suggest a pathway to tunable mechanics in adaptive metamaterials, with implications for mechanical regulation in biological systems.

\end{abstract}

\maketitle


%

\section{Introduction} 

{
The study of active solids~\cite{Xu2023, Tan2022, Baconnier2022, briand2018spontaneously, Scheibner2020, Banerjee2021} is inspired by living systems such as  tissues and cell assemblies~\cite{Szabo2006, Bi2015, henkes2020dense}, bacterial biofilms~\cite{billings2015material, Shankar2021}, and artificial material such as smart matter~\cite{Nguyen2018, Vernerey2019}.  These systems are inherently out of equilibrium at the microscopic scale: their constituents locally consume and dissipate energy, generating stresses or propulsion and breaking detailed balance. Unlike crystalline solids, which break continuous translational and rotational symmetries in equilibrium, amorphous solids are stabilized by compression and by local force and torque balance \cite{Aranson2006,Falk1998, Jaeger1996, Baule2018}. In active solids, maintaining this balance is essential for mechanical stability;  its breakdown naturally leads to deformation and flow.

Inspired by biological motility, artificial micro- and nano-swimmers have been developed that convert ambient energy into persistent motion and are well described by active Brownian particle dynamics~\cite{bechinger2016active, marchetti2013hydrodynamics}. At high densities, such activity can profoundly alter the mechanical properties of solids. Recent studies have reported extreme deformations in active crystals~\cite{Shi2023} and yielding in active amorphous solids~\cite{Goswami2025}.  In equilibrium solids, low-energy deformations are captured by elastic displacement fields, while plasticity and yielding arise from defect-mediated, singular distortions~\cite{sethna2017deformation, anderson2017theory, phillips2001crystals, Ozawa2018}.  In amorphous solids, however, the absence of broken translational symmetry obscures this distinction, and localized, non-singular rearrangements play a significant role~\cite{Zaccone_2014, falk2011def, Ghosh_2017}. While homogeneous elasticity predicts a linear stress–strain relation, this description breaks down in the presence of disorder or spatially heterogeneous activity~\cite{DiDonna2005}.

A natural framework to capture such effects is the decomposition of particle displacements into affine and nonaffine components~\cite{Falk1998, Das2010, ganguly2013nonaffine}. The affine part describes uniform elastic distortions, whereas the nonaffine part encodes residual, disorder-driven rearrangements and defect precursors. In active solids, self-propulsion acts as a persistent, microscopic source of stress that enhances nonaffine motion even in the absence of external loading, making activity-induced nonaffinity a key ingredient in the mechanical response.

We investigate spontaneous deformations in active solids by quantifying local non-affine fluctuations relative to the global strain. Our main achievements are threefold. First, using scaling arguments and numerical simulations, we predict how the steady-state mean non-affinity depends on activity parameters -- active speed and persistence -- revealing behavior beyond simple active-temperature estimates. Second, at higher activity, pronounced non-affine zones coexist with regions of small non-affinity, preceding the divergence of non-affinity, defect proliferation, and eventual melting. Correlation functions show that the non-affine correlation length increases with active speed and grows to finally saturate with active persistence, while the relaxation of system-averaged non-affinity is set by the active persistence time. Third, we demonstrate that spatially patterned activation provides a direct route to locally induce non-affinity and mechanical softening. Together, these results establish a quantitative framework for understanding and controlling the interplay between activity, non-affine deformations, and mechanical stability in dense active materials.

The paper is organized as follows. In Sec.~\ref{model}, we introduce the model of the active solid. Sec.~\ref{non_affinity} presents the calculation of non-affine fluctuations, the steady-state scaling relations, and the local distributions and correlations of non-affinity. Sec.~\ref{structural_analysis} examines changes in structural indicators, including the shear modulus, solid and hexatic order parameters, and topological defect proliferation. Sec.~\ref{control} demonstrates how spatially patterned activity can locally induce non-affinity and soften the solid. Finally, Sec.~\ref{conc} summarizes the main results and provides a discussion and outlook.
}

\section{Model}
\label{model}
We consider an excluded volume solid in two dimensions characterized by a triangular lattice and quasi-long-range order, composed of $N = n_x n_y$ particles interacting via the Weeks-Chandler-Andersen (WCA) potential~\cite{Weeks1971, Chandler1983}, $U(r_{ij}) = 4\epsilon [ {(\sigma/r_{ij})^{12}}-(\sigma/r_{ij})^6] +\epsilon$ for $|r_{ij}| < r_c = 2^{1/6} \s$ and $U(r_{ij}) =0$ otherwise. The system is maintained at high density $\rho = N / L_x L_y$ with mean inter-particle separation $a = [{2}/{\sqrt{3} \rho}]^{1/2}$.
The equilibrium phase diagram, based on the WCA potential, reveals that for fixed temperature $k_B T / \varepsilon = 1.0$, the solid undergoes a continuous BKTHNY transition to a hexatic phase at $\rho \sigma^2 = 0.92$, followed by first-order melting to a liquid at $\rho \sigma^2 = 0.906$~\cite{Khali2021}.

{
Modeling the constituents of this high-density solid ($\rho \sigma^2 \geq 1$)  as Active Brownian Particles (ABPs), we investigate its properties under increasing activity, controlled by the active speed and persistence.
}
Each particle moves with self-propulsion speed $v_0$ in a direction $\hat{\mathbf{n}}_i = {\cos \theta_i(t), \sin \theta_i(t)}$, with the angle $\theta_i(t)$ evolving via rotational diffusion. The dynamics of the $i$-th particle is governed by the following coupled Langevin equations:
\begin{align}
    d\rv_i(t) &= v_0\hat{\bf n}_i(t) dt - \mu \nabla_i\sum_{j \epsilon R_i} U(r_{ij}) dt + \sqrt{2D_t}\, d{\bf B}_i(t) \nn \\
    d{\theta_i}(t) &=  \sqrt{2D_r}\,dB^{r}_i(t), 
    \label{eq_Lange}
\end{align} 
where $D_t$ and $D_r$ are the translational and rotational diffusivities, and $d\mathbf{B}_i(t)$ and $dB_i^r(t)$ are independent Gaussian white noise terms with mean zero and variance $dt$. The Euler-Maruyama method is used to directly integrate the dimensionless forms of the above equations (see Appendix-\ref{app_dim}) in numerical simulations. The units of energy, length, and time are set by $\epsilon$, $\sigma$, $\t_u = \s^2/\mu \e$. 
The system property is controlled by the following dimensionless parameters: active speed $\L = v_0 \s/\mu \e$, translational diffusivity $\tilde D_t = D_t/\mu \e$, and orientational diffusivity $\tilde D_r= D_r \s^2/\mu \e$. Unless stated otherwise, we set $\tilde D_t = 1$ for thermal systems and $\tilde D_t = 0$ for athermal systems, and fix the density at $\rho \sigma^2 = 1.1$. We examine the effects of activity by varying ${\L}$ and $\tilde D_r$, and additionally investigate the effect of changing density $\rho \sigma^2$ in one case.


\section{Non-affine parameter}  
\label{non_affinity}

To characterize deviations from uniform, affine deformations in active solids, we define a non-affinity parameter, which measures the extent of local rearrangements relative to a global strain. Specifically, for a given local region, it quantifies the discrepancy between the observed particle displacements and those predicted by a best-fit affine transformation. 

\begin{figure}[t!] 
\centering 
\includegraphics[width=0.99\linewidth]{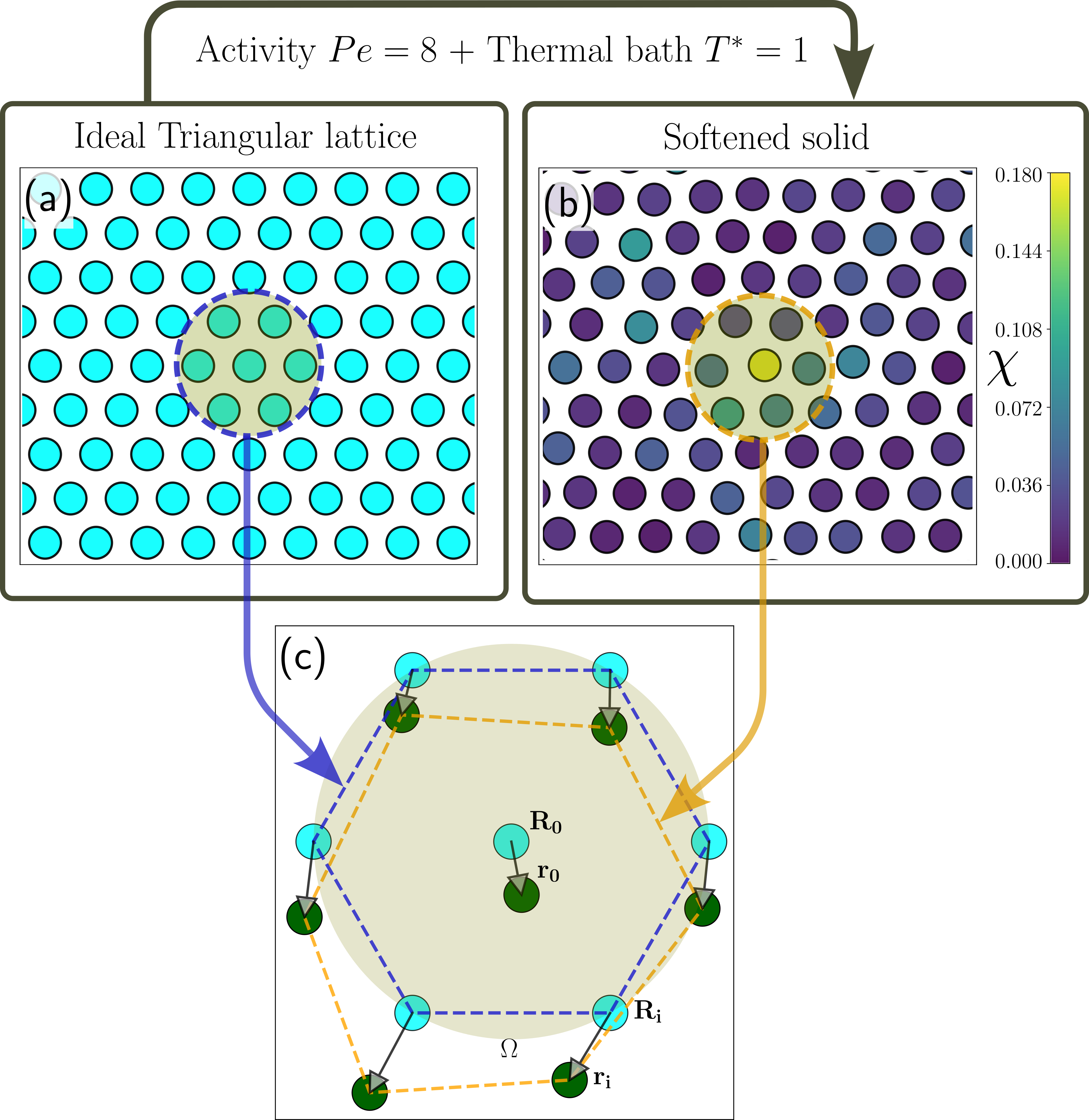} 
.\caption{
A schematic illustration of the activated solid. (a)~An ideal triangular lattice that serves as the reference for the calculation of non-affinity $\chi$. (b)~A representative configuration at $\tilde D_t = 1.0$ and $\Lambda = 8.0$, is shown color coded by local non-affine parameter $\chi$; see the color bar. 
(c)~A single coarse-graining region $\Omega$ is defined as the volume enclosed by the first nearest neighbors, determined by mean inter-particle separation $a$ at a given density $\r$. This region includes one central particle with reference position $\mathbf{R}_0$~(cyan) and actual location $\mathbf{r}_0$~(dark green). The reference~(cyan, $\{\mathbf{R}_i\}$) and actual locations~(dark green, $\{\mathbf{r}_i\}$) of its six nearest neighbors are shown as well. 
}
\label{fig:schematic}
\end{figure}

The non-affine fluctuation $\chi_0$ of the $0$-th particle, at lattice position ${\mathbf{R}_0}$, is defined within a coarse-graining volume $\Omega$ containing all six nearest neighbors $\{\mathbf{R}_i\}$ of ${\mathbf{R}_0}$ (see Fig.~\ref{fig:schematic}c). Instantaneous positions ${\mathbf{r}_0}$ and $\{\mathbf{r}_i\}$ yield displacements $\mathbf{u}_0 = \mathbf{r}_0 - \mathbf{R}_0$ and $\mathbf{u}_i = \mathbf{r}_i - \mathbf{R}_i$ relative to reference coordinates $\{\mathbf{R}_i\}$ on an ideal triangular lattice.

The optimal estimate of local strain matrix ${\cal E}$ is obtained by minimizing the mean-square difference $D^2$ between the observed relative displacements $\mathbf{\Delta}_i = \mathbf{u}_i - \mathbf{u}_0$ and estimated displacements ${\cal E} [\mathbf{R}_i - \mathbf{R}_0]$~\cite{Falk1998}, 
\begin{equation}
D^2
= \sum_i \left( \mathbf{\Delta}_i - {\cal E}[\mathbf{R}_i-\mathbf{R}_0] \right)^2. \label{eq:chi} 
\end{equation}
Note that $D$ has the dimension of length. After minimizing $D^2$ at a given density and activity, one obtains the $\alpha \beta$-th component of the optimal strain matrix, the affine strain,  
\begin{equation}
{\cal E}^m_{\alpha \beta} = \sum_{\gamma} x_{\alpha \gamma } y_{\beta \gamma}^{-1} - \delta_{\alpha\beta}, 
\label{eq_Dab}
\end{equation}
where 
\begin{align}
  x_{\alpha \beta} &= \sum_{i=1}^{N_{\Omega}} \left( r_{i \alpha} - r_{0 \alpha} \right) \left( R_{i \beta} - R_{0 \beta} \right), \nonumber\\
  y_{\alpha \beta} &= \sum_{i=1}^{N_{\Omega}} \left( R_{i \alpha} - R_{0 \alpha} \right) \left( R_{i \beta} - R_{0 \beta} \right). \nn
\end{align}

The residual $D^2$ value using the minimized ${\cal E}$ denoted by ${\cal E}^m$ in Eq.\eqref{eq_Dab} gives the non-affine fluctuation $\chi_0$ of the $0$-th particle~\cite{Falk1998, Das2010}:
 \bea
 \chi_0 = \sum_i \left( \mathbf{\Delta}_i - {\cal E}^m[\mathbf{R}_i-\mathbf{R}_0] \right)^2.
 \eea
{ Applying this method to all particles results in the local non-affinity $\chi_i$ for each particle $i$ in the system.
A projection operator formalism showed that this procedure projects the relative displacements $ \mathbf{\Delta}_i$ onto mutually orthogonal subspaces of the affine and non-affine fluctuations, characterized by the affine strain tensor ${\cal E}^m$ and the scalar non-affine parameter $\chi_i$~\cite{ganguly2013nonaffine}.}

The global non-affine parameter $X = N^{-1} \sum_{i=1}^{N} \chi_i$ is the  system average with $\chi_i$ denoting non-affine parameter associated with $i$-th particle.  Note that $\chi_i$ and $X$ have dimensions of length squared, which we use in the next section to develop a scaling argument.

\subsection{Scaling analysis}

{
The average total non-affinity, $\langle \sum_i \chi_i \rangle = \mathrm{Tr}[P C]$~\cite{ganguly2013nonaffine}, with $C = \langle \Delta \Delta^T \rangle$ and $\Delta_i = {\bf u}_i - {\bf u}_0$, connects non-affinity to relative displacements. The projection operators $P = I - Q$  and $Q = R(R^T R)^{-1} R^T$ are symmetric and idempotent projection operators, projecting onto non-affine and affine subspaces, respectively, with $R$ a block matrix of identities and zeros determined by lattice geometry. Consequently, the mean global non-affinity, $\langle X \rangle$, scales with $\langle C \rangle \sim \ell^2$, directly linking non-affinity to $\ell^2$, the mean-squared displacement of particles~\footnote{More precisely, $\ell^2$ corresponds to the mean-squared displacement of the normal modes}.
}

{
In an equilibrium solid with harmonic interactions of stiffness $k$, the single-particle diffusivity $D_t$ and the harmonic relaxation rate $\mu k$ set the mean-squared displacement, $\ell_{\rm eq}^2 \sim D_t/(\mu k)$.}
For an equilibrium harmonic solid, the $n$-th moment of $X$, was shown to scale as $\langle X^n \rangle_{eq} \sim \ell_{eq}^{2n}$~\cite{ganguly2013nonaffine}. 
In sterically stabilized solids, $ k $ is determined by $ U''(a) $, and therefore increases with density. 
{
For a two-dimensional isotropic solid, the shear and Young’s moduli scale linearly with $k$ and, more generally, increase nonlinearly with density~\cite{Sengupta2000}}. This elastic stiffening implies a corresponding decrease in the mean non-affinity, $ \langle X \rangle $. Approaching melting, however, the shear modulus ${\cal G}$ vanishes, and the non-affine parameter is expected to diverge. These considerations motivate a generalized expression for equilibrium non-affine rearrangements, $\langle X \rangle_{eq} \sim \frac{D_t}{{\cal G}}$, with ${\cal G}$ increasing with density $\r$ and vanishing at the melting transition.  Ignoring non-linearities for simplicity, the density dependence of shear modulus can be approximated as:
\bea
{\cal G}(\rho) & \propto & (\rho - \rho_m) \quad \text{for} \quad \rho > \rho_m, \nn\\  
&=& 0 \quad \text{otherwise},
\label{eq_G1}
\eea
with $ \rho_m $ representing the melting density.
As shown below, this hypothesis predicts that non-affinity decreases with increasing density, a trend confirmed by numerical simulations.

In this paper, we consider the impact of activity on a dense solid. 
{One might expect that replacing the thermal diffusivity $D_t$ with the effective active diffusivity $v_0^2/2 D_r$~\cite{Howse2007, Shee2020} in $D_t/{\cal G}$ would account for the activity-induced excess non-affinity.}
However, this simple substitution fails in a solid, where each test particle is trapped within a cage formed by its neighbors. Approximating this confinement by a harmonic potential with stiffness $k$, we can invoke an earlier estimate for the steady-state mean-squared displacement arising from activity $\ell_{\rm ac}^2 = v_0^2/[\mu k (D_r + \mu k)]$, as obtained in Ref.~\cite{Chaudhuri2021}.
This leads us to hypothesize that activity induces an excess non-affinity of the form
$\langle X^{(a)} \rangle \approx v_0^2/[\a {\cal G } (D_r + \a {\cal G})]$, 
where the spring constant is replaced by the more appropriate solid descriptor, the shear modulus ${\cal G}$; here $\alpha$ is a proportionality constant. 
Thus, the total non-affine parameter can be written in the following dimensionless form as
{
\bea
\f{\la X \ra}{\s^2} \approx \f{\tilde D_t }{\a \tilde{\cal G}} +\f{\Lambda^2}{\a {\tilde{\cal G}}(\tilde D_r + \a \tilde{\cal G} )} .
\eea
}
{In the above expression, we used dimensionless forms of translational diffusivity $\tilde D_t = D_t/\mu \e$, shear modulus $ \tilde{\cal G} = {\cal G} \s^2/\e$, rotational diffusivity $\tilde D_r = D_r \s^2/\mu\e$ and active speed $\Lambda = v_0 \s/\mu\e$.}

In active systems, the melting point $\r_m$ depends on $\Lambda$~\cite{Paliwal2020}, while otherwise preserving the same density dependence of the shear modulus ${\cal G}$ as in Eq.\eqref{eq_G1}. Beyond this shift in melting point, increasing activity leads to a reduction of the shear modulus according to
\bea 
{\cal G} = {\cal G}_0 - {\cal G}_1 \Lambda^2,
\label{eq_G2}
\eea
as demonstrated in a later section through numerical simulations (Fig.~\ref{fig:struct}). 
However, 
{a small value of ${\cal G}_1$ ensures that} the $\Lambda$ dependence of ${\cal G}$ does not modify the dominant pre-melting scaling of the 
excess non-affinity due to activity $\la X^{(a)} \ra =  \la X \ra -  \la X \ra_{eq} \sim \Lambda^2$ appreciably.

\begin{figure*}[t!] 
\centering 
\includegraphics[width=0.95\textwidth]{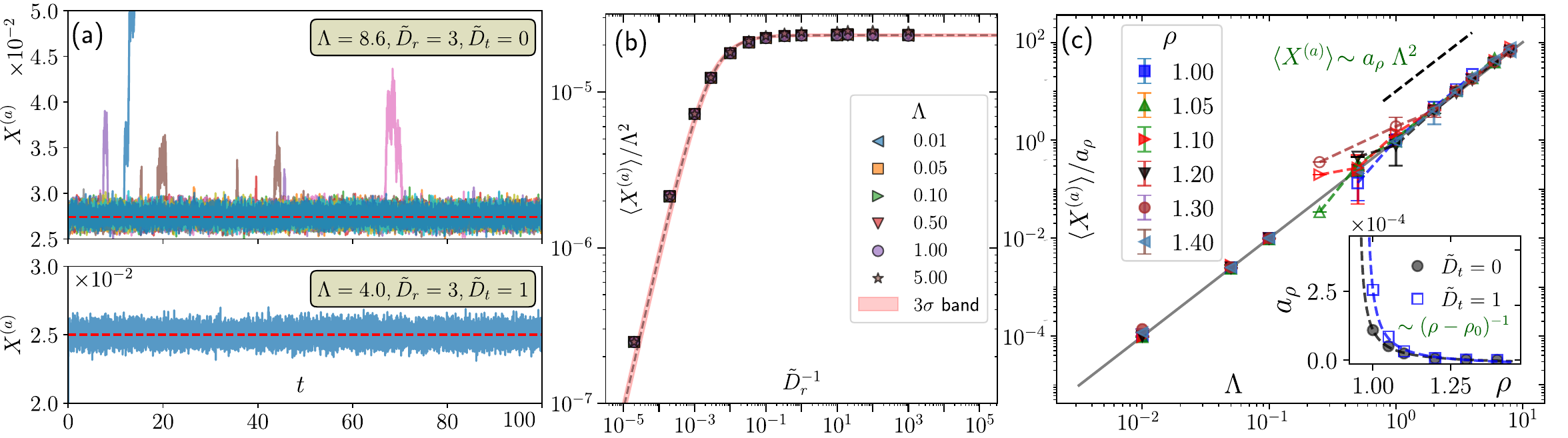} 
\caption{
{
(a) Representative time series of the system-averaged non-affinity $X(t)$ at $\Lambda = 8.6$ (athermal, top) and $\Lambda = 4.0$ (thermal, bottom); other parameters are given in the legends.
(b) Scaled excess non-affinity $\langle X^{(a)} \rangle / \Lambda^2$ as a function of the active persistence $\tilde D_r^{-1}$. The data are well fit by $\langle X^{(a)} \rangle / \Lambda^2 = a/(D_r + c)$, with $a = 9.21 \times 10^{-3} \pm 4.57 \times 10^{-4}$ and $c = 397.96 \pm 20.57$. At large $\tilde D_r$ (low persistence), $\langle X^{(a)} \rangle$ grows approximately linearly with the persistence time $\tilde D_r^{-1}$ and saturates beyond $\tilde D_r^{-1} > c^{-1}$.
(c) $\langle X^{(a)} \rangle / a_{\rho}$ versus $\Lambda$ for $N = 4096$ particles at fixed $\tilde D_r=3.0$ and different densities $\rho$, where $X^{(a)} = X(\Lambda) - X_{\mathrm{eq}}$. Filled (open) symbols correspond to athermal ($\tilde D_t = 0$) and thermal ($\tilde D_t = 1.0$) systems, respectively. Fits to $a_{\rho}\Lambda^{\nu}$ yield $\nu = 2.00 \pm 0.01$ (athermal) and $\nu = 2.11 \pm 0.06$ (thermal). Inset: density dependence of $a_{\rho} \sim (\rho - \rho_0)^{-1}$, with $\rho_0 \sigma^2 = 0.947 \pm 0.008$ (athermal) and $0.971 \pm 0.004$ (thermal).}
}
\label{fig:scaling_X}
\end{figure*}

In summary, for densities $\r > \r_m$ and active speed $\Lambda < \Lambda_c$ allowing the system to remain in the solid phase, the following scaling form for $ \la X^{(a)} \ra$ is expected:
\bea
\la X^{(a)} \ra \sim \f{\Lambda^2}{\tilde{\cal G} [\tilde D_r + \a \tilde{\cal G}]  }. 
\label{eq:XscaleAll}
\eea
We compare these predictions with numerical simulations below.

\subsection{Scaling verification via numerical simulations}
\label{subsec:scaling_X_sim}
For this purpose, we analyze the system-averaged non-affinity $X = N^{-1} \sum_i^N \chi_i$. 
At $\rho \sigma^2 = 1.1$ and $\tilde D_r = 3$, the time evolution $X(t)$ reaches a steady state for $\Lambda \lesssim 8$, whereas for $\Lambda \gtrsim 8$ it may diverge due to irreversible reorganization of particle positions. 
{
As we show later, the melting of the solid associated with the proliferation of topological defects occurs at a larger active speed, $\Lambda = 10$; indicating that the divergence of non-affinity precedes solid melting, in agreement with earlier observations in Ref.~\cite{popli2019exploring}. 

Representative trajectories $X(t)$ at $\Lambda = 4$ ($\tilde D_t = 1$) and $\Lambda = 8.6$ ($\tilde D_t = 0$) are shown in Fig.~\ref{fig:scaling_X}(a).
At high active speeds, the trajectories of $X$ exhibit large excursions that can subsequently relax back toward baseline values, indicating the healing of nascent plastic deformations. In contrast, as illustrated in the top panel of Fig.~\ref{fig:scaling_X}(a), some trajectories undergo runaway growth, signaling irreversible plastic flow. In the following, we therefore restrict our analysis to the steady-state regime $\Lambda \lesssim 8$ to test the scaling hypothesis proposed above.
}

We now consider the excess non-affinity induced by activity, $ X^{(a)}(\Lambda,\rho,T) = X(\Lambda,\tilde D_r, \rho,T) - X_{eq}(\rho,T)$.  
{This quantity can be obtained in two equivalent ways. One approach is to simulate an active system coupled to a thermal bath ($\tilde D_t = 1$) and compare it with the corresponding equilibrium system at $\Lambda = 0$, thereby determining $X(\Lambda,\tilde D_r,\rho,T)$ and $X_{\mathrm{eq}}(\rho,T)$ separately. Alternatively, one may perform athermal simulations with $\tilde D_t = 0$, which directly isolate the contribution of activity to the system's non-affinity. 

This second approach is adopted in the numerical simulations shown in Fig.~\ref{fig:scaling_X}(b), which use $N = 4096$ ABPs at $\rho \sigma^2 = 1.1$. With increasing persistence, the mean non-affinity initially grows linearly with the persistence time $\tilde D_r^{-1}$ and subsequently saturates, in agreement with the scaling prediction of Eq.~\eqref{eq:XscaleAll}. This behavior persists over the entire range of $\Lambda$ for which $X$ attains a steady state. A clear data collapse is obtained when plotting $\langle X \rangle/\Lambda^2$, which is well described by a nonlinear fit of the form $a/(D_r + c)$, with fit parameters $a = 9.21\times10^{-3} \pm 4.57\times10^{-4}$ 
and $c = 397.96 \pm 20.57$. }

In Fig.~\ref{fig:scaling_X}(c), we present simulations of $N=4096$ ABPs for $\tilde D_r=3.0$, $\Lambda \in [0.5,8.0]$, densities $\rho \in [1.0,1.4]$, and $\tilde D_t = 1.0$, from which we extract the excess steady-state non-affinity due to activity $\la X^{(a)}\ra$. These results are compared with athermal simulations performed at $\tilde D_t = 0$. A reasonable data collapse is obtained when results at different densities are plotted as $\langle X^{(a)} \rangle / a_{\rho}$ versus $\Lambda$.
The excess non-affinity follows a scaling form $\langle X^{(a)} \rangle / a_{\rho} \sim \Lambda^{\nu}$ with exponent $\nu \approx 2.0$, as shown in Fig.~\ref{fig:scaling_X}(c). 

{At low persistence, Eq.~\eqref{eq:XscaleAll} predicts $\langle X^{(a)} \rangle \sim 1/\tilde{\cal G} \sim (\rho - \rho_0)^{-1}$, while at high persistence it scales as $\langle X^{(a)} \rangle \sim 1/\tilde{\cal G}^2 \sim (\rho - \rho_0)^{-2}$, where $\rho_0$ denotes the melting density in active solid. In the current regime ($\tilde D_r = 3.0$), $\langle X^{(a)} \rangle \sim \tilde D_r^{-1}$ is consistent with the low-persistence scaling. The density-dependent prefactor, $a_\rho \sim (\rho - \rho_0)^{-1}$, shown in the inset of Fig.~\ref{fig:scaling_X}(c), follows this scaling, with the fitted value $\rho_0 \sigma^2 = 0.971 \pm 0.004$ exceeding the equilibrium melting density $\rho_m^{\mathrm{eq}} \sigma^2 = 0.92$~\cite{Khali2020}, reflecting activity-induced softening of the solid.}

Note that for $\Lambda \lesssim 1$, thermal fluctuations in simulations with $\tilde D_t = 1$ dominate over activity, rendering the $\Lambda$-dependence of the non-affinity difficult to resolve. Results from athermal simulations ($\tilde D_t = 0$) over the range $\Lambda \in [10^{-2}, 8]$ are therefore shown using filled symbols. These data collapse onto a single curve obeying $\langle X^{(a)} \rangle \sim \Lambda^2$ across nearly three decades in $\Lambda$ and for a wide range of densities.
The density dependence, encapsulated by $a_{\rho} \sim (\rho - \rho_0)^{-1}$ with $\rho_0 \sigma^2 = 0.947 \pm 0.008$ (inset of Fig.~\ref{fig:scaling_X}(c)), yields a lower fitted value of $\rho_0 \sigma^2$, indicating a reduced melting density in the absence of thermal noise, as expected. Together, these results confirm both scaling forms predicted in Eq.~\eqref{eq:XscaleAll}. The robustness of the observed scaling behavior is further corroborated by simulations of larger system sizes, as discussed in Appendix~\ref{app_scaling}.

{
The scaling of global non-affinity with activity parameters reflects the mechanisms by which activity drives particle rearrangements in solids. Increasing the self-propulsion speed, $\Lambda$, strengthens active fluctuations and thereby enhances non-affinity. Increasing the persistence time, $\tilde D_r^{-1}$, allows particles to keep their heading directions unchanged for longer durations, promoting more coherent local rearrangements. Consequently, $\langle X \rangle$ increases with persistence at low to moderate values, but saturates at large persistence as the system gets effectively jammed. This behavior arises from the competition between persistent active forcing and structural constraints of the solid, and contrasts with equilibrium solids, where particle displacements are isotropic, homogeneous, and Gaussian.
}

%
{
\subsection{Coexistence}

In Fig.\ref{fig:Pchi}, we show the steady-state probability distributions $P(\chi)$ of local non-affinity $\chi_i=\chi$ for an athermal system ($\tilde D_t=0$) on log-log axes for varying active speed (panel (a)\,) and persistence time (panel (b)\,). In equilibrium, the distribution is described by a chi-squared form~\cite{ganguly2015statistics}, 
{
reflecting the Gaussian statistics of particle displacements and the quadratic dependence of the local non-affinity on these fluctuations.} 
{We observe a clear departure from this expectation in the spontaneous fluctuations of active solids, consistent with persistence rendering non-Gaussianity to displacement fluctuations~\cite{Malakar2020}. } As the activity parameters increase, the slightly skewed distributions broaden, get more skewed, and develop a heavy tail, followed by the appearance of a secondary maximum indicating rare but large non-affine rearrangements.  This bimodality, arising from pronounced non-affine zones within an otherwise normal solid, precedes the divergence of non-affinity and serves as a precursor to defect formation and eventual melting. 
{The statistical weight of large non-affinity originates from either transient formation and decay of highly non-affine regions or steady-state coexistence with local plastic zones; both types of events are illustrated in Appendix-\ref{sec_conf_coex} using particle configurations and time series of the global non-affinity.}

\begin{figure}[t!]
    \centering
    \includegraphics[width=0.99\linewidth]{{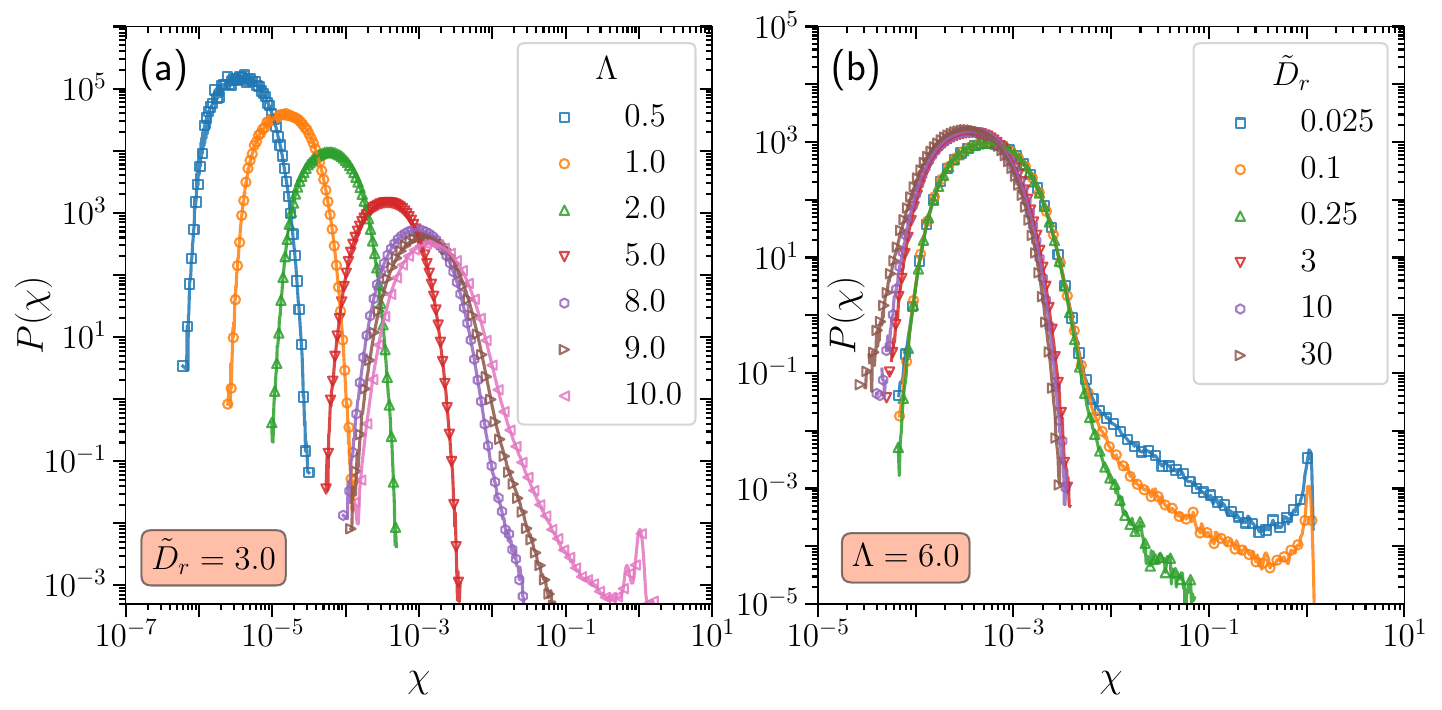}} 
    \caption{
    {
    Probability distributions $P(\chi)$ of local non-affinity $\chi_i=\chi$ on log-log axes. (a) At fixed $\tilde D_r=3.0$, increasing active speed $\Lambda$ shifts and broadens the distributions toward larger $\chi$, with a secondary high-$\chi$ peak emerging at large $\Lambda$. (b) At fixed $\Lambda=6.0$, decreasing $\tilde D_r$ drives a crossover from unimodal to bimodal distributions, reflecting the onset of rare, large non-affine rearrangements. 
      }
 }
    \label{fig:Pchi}
\end{figure}

In panel (a), at fixed rotational diffusion $\tilde D_r = 3.0$, increasing the self-propulsion speed $\L$ systematically shifts the skewed distribution toward larger values of $\chi$ and broadens it, indicating enhanced non-affine rearrangements consistent with the increase of the mean $\la X \ra \sim \L^2$. At sufficiently high $\L$, the distribution becomes more skewed, first developing a heavy tail at large $\chi$ and then a secondary peak, signaling the coexistence of dominant small non-affine fluctuations with rare large rearrangements. 

Fig.~\ref{fig:Pchi}(b) shows the effect of varying $\tilde D_r$ at fixed active speed $\L=6.0$. For small persistence ($\tilde D_r>0.25$), the distributions remain unimodal; the low-$\chi$ tail is progressively suppressed, shifting probability weight toward larger $\chi$ as persistence increases ($\tilde D_r$ decreases). This behavior is consistent with the linear increase of $\langle X\rangle$ in the small-persistence regime. At $\tilde D_r=0.25$, the distribution undergoes a qualitative change, marked by a clear rightward shift and broadening of the peak. For $\tilde D_r \leq 0.25$, the main maximum -- carrying the bulk of the probability -- saturates and becomes insensitive to further increases in persistence. A long tail at large $\chi$ develops for $\tilde D_r \leq 0.25$, eventually giving rise to a secondary peak at high $\chi$, rendering the distribution bimodal. This reflects the coexistence of dominant small non-affine fluctuations with rare but substantial non-affine rearrangements.

Further, the change in the character of the probability distributions $P(\chi)$ across a broad range of $\Lambda$ and $\tilde D_r$, highlighting the emergence of coexistence, is shown in Appendix~\ref{app_Pchi}. This detailed analysis reveals that increasing active speed broadens and shifts $P(\chi)$ toward larger values, while intermediate persistence promotes bimodality, reflecting the coexistence of small and rare large non-affine rearrangements, with very high or very low persistence restoring unimodal, tail-suppressed distributions.
}

\begin{figure}[t!]
    \centering
    \includegraphics[width=0.99\linewidth]{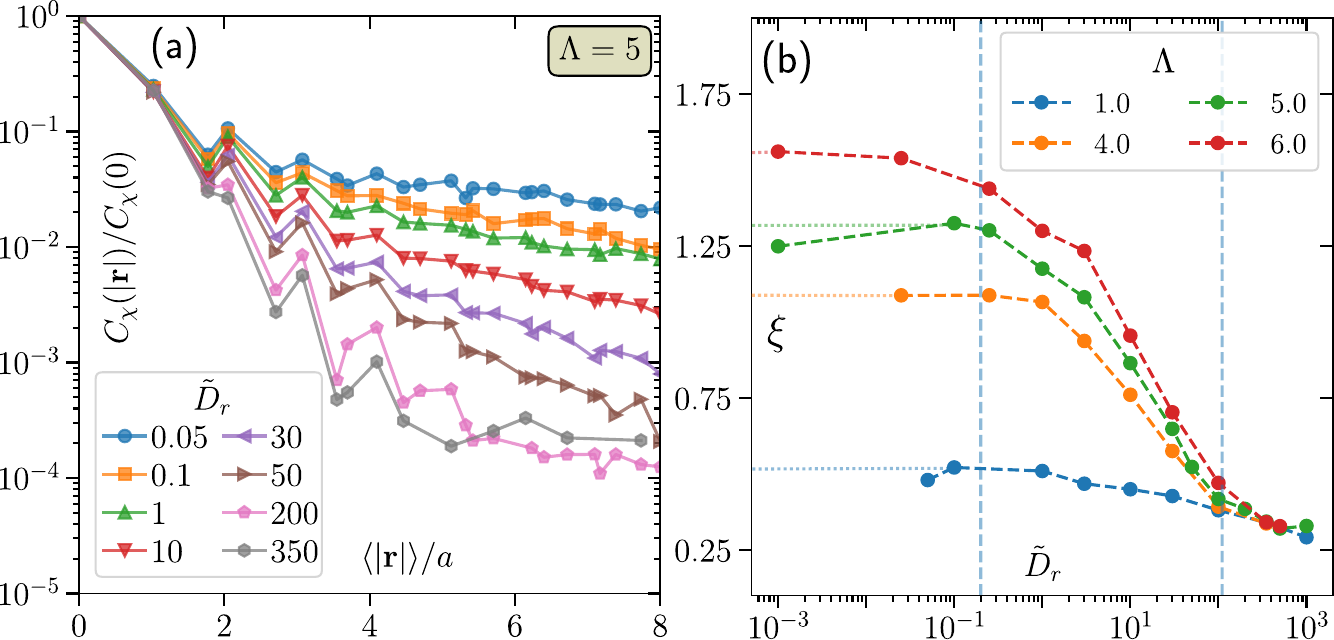}
    \caption{
        Normalized spatial correlations of the local non-affinity $\chi$ (a) at $\Lambda=5$ for varying $\tilde D_r$ and (b) corresponding correlation length $\xi$ across active speeds $\L$ and rotational diffusivities $\tilde D_r$. $\xi$ increases with active speed and persistence in the persistent-active and intermediate regimes, while in the low-persistence limit, correlations are short-ranged and largely $\L$-independent.
    }
    \label{fig:xi_OZ}
\end{figure}

{
\subsection{Correlations}
To further characterize the local properties of non-affinity,  we analyze the spatial correlation 
\bea
C_\chi(r) = \la \chi_i \chi_j \, \d(r_{ij} -r) \ra
\eea
at steady state, where $r = |\rv|$, $\chi_i$ denotes the local non-affinity for particle $i$, and $r_{ij}=|\rv_i - \rv_j|$ is the distance  between particles $i$ and $j$.  
The normalized two-point correlation function $C_\chi(|\mathbf{r}|)/C_\chi(0)$ at fixed activity $\Lambda=5$ is shown in Fig.~\ref{fig:xi_OZ}(a). These correlations clearly extend over longer distances as persistence increases (i.e., as $\tilde D_r^{-1}$ grows). We extract a characteristic correlation length $\xi$ using the integral definition $\xi = \int dr\, C_\chi(r)/C_\chi(0)$, which captures the correct trends, although the precise numerical value can vary depending on the specific choice of correlation length measure.
Mapping $\xi$ across a range of $\Lambda$ and $\tilde D_r$ (Fig.~\ref{fig:xi_OZ}(b)) reveals three distinct regimes. For high persistence, $\tilde D_r \lesssim 0.2$, $\xi$ saturates to a $\Lambda$-dependent plateau, reflecting extended correlations in persistent-active systems. In the intermediate regime, $0.2 \lesssim \tilde D_r \lesssim 110$, $\xi$ decreases monotonically with $\tilde D_r$, indicating progressively shorter-ranged correlations as persistence weakens. Finally, in the low-persistence regime, $\tilde D_r \gtrsim 110$, the short correlation lengths become largely independent of activity. Increasing $\Lambda$ systematically enhances $\xi$ in the persistent-active and intermediate regimes, showing that stronger self-propulsion promotes correlations of non-affine rearrangements over longer distances, whereas in the low-persistence limit the correlation length collapses to a small value, independent of $\L$.
}

\begin{figure}[ht]
    \centering
    \includegraphics[width=0.5\linewidth]{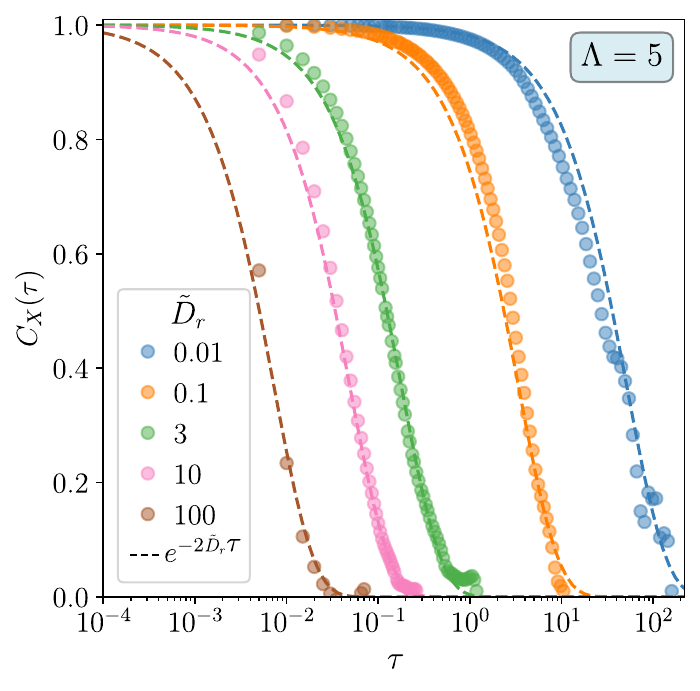}
    \caption{
        Normalized time autocorrelation $C_X(\tau)$ at $\Lambda=5$ for several rotational diffusion constants $\tilde D_r$. Dashed lines indicate the reference exponential decay $\exp(-2 \tilde D_r \tau)$. }
    \label{fig:Xtcorr}
\end{figure}

{
In addition, we compute the two-time autocorrelation of the system-averaged non-affinity $X$, using the normalized definition
\bea
C_{X}(\tau)=\langle \delta X^{(a)}(t)\,\delta X^{(a)}(t+\tau)\rangle/\langle [\delta X^{(a)}(t)]^2\rangle
\eea
where $\delta X^{(a)}(t) = X^{(a)}(t) - \langle X^{(a)}(t) \rangle$ and the averaging is performed over time $t$. Our simulations show that $C_X(\tau)$ decays as $C_X(\tau) \sim \exp(-2 \tilde D_r \tau)$, provided the system has reached a steady state in terms of non-affine fluctuations. As an example, in Fig.~\ref{fig:Xtcorr} we show numerical estimates of $C_X(\tau)$ at $\Lambda=5$ confirming this exponential decay for all tested values of $\tilde D_r$. This demonstrates that the dynamics of non-affinity under activity are primarily governed by the persistence of the active motion.
}

{
Active solids display behavior that is fundamentally different from their equilibrium counterparts. In equilibrium systems, particle displacements are governed by Gaussian fluctuations, leading to chi-squared distributions of local non-affinity and correlations that decay within roughly two lattice spacings~\cite{ganguly2015statistics, popli2018translationally}. Active particles, however, apply persistently directed forces along their instantaneous headings. While at low persistence they behave similarly to passive particles, increasing persistence enhances directional, non-Gaussian fluctuations. As a result, the distributions of local non-affinity deviate strongly from the chi-squared form (Fig.~\ref{fig:Pchi}), and the associated correlation length grows with persistence (Fig.~\ref{fig:xi_OZ}), indicating the emergence of longer-ranged cooperative rearrangements absent in equilibrium solids.

}

\section{Analyzing structural indicators}
\label{structural_analysis}
Thus far, we have characterized the softening of a high-density solid through the non-affine parameter $X$ and its scaling with activity characterized by persistence and active speed, and density. In this section, we extend our analysis to include elastic and structural indicators, such as the shear modulus, solid order, and hexatic order, to further explore the active softening of the solid.  
We perform this analysis using $\rho \sigma^2 = 1.1$, $\tilde D_r = 3.0$ and $\tilde D_t = 1.0$.

\begin{figure*}[t!] 
\centering
\includegraphics[width=0.95\textwidth]{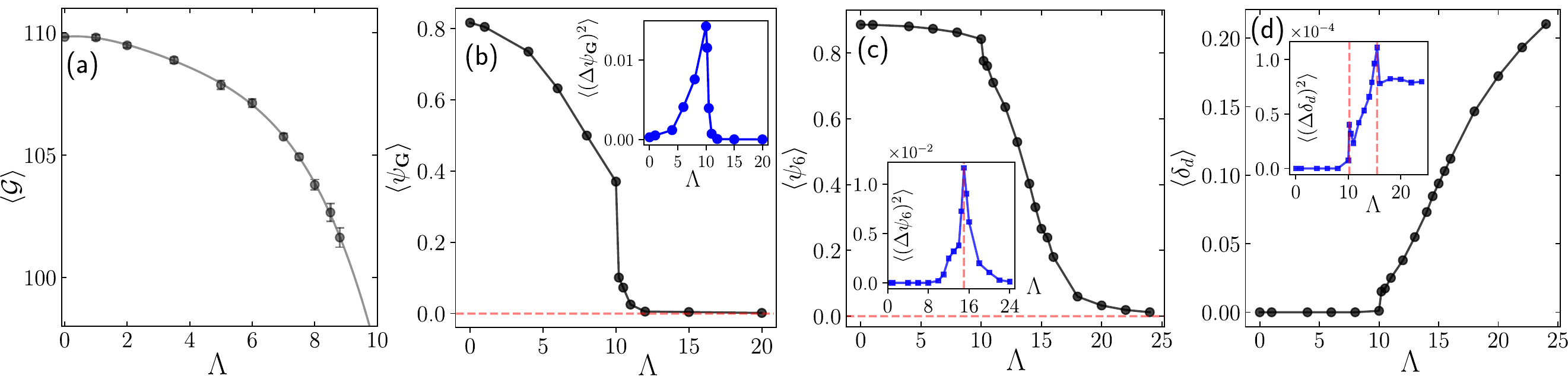}   
\caption{ 
(a): Shear modulus $ \langle \mathcal{G} \rangle $ averaged over $ n_{\mathrm{lin}} $ stress–strain curves showing linear response, plotted against $ \Lambda $ for a system at density $ \rho \sigma^2 = 1.1 $. 
(b), (c), (d): Variation of solid order $ \langle \psi_{\mathbf G} \rangle $, hexatic order $ \langle \psi_6 \rangle $, and defect fraction $ \langle \delta_d \rangle $ with $ \Lambda $;  respective variances are shown in the insets. $\langle (\Delta \psi_\mathbf{G})^2 \rangle$ peaks at $\Lambda=10.0$, $\langle (\Delta \psi_6)^2 \rangle$ shows a peak at $\Lambda = 15.0$. Whereas $\langle (\Delta \delta_d)^2 \rangle$ exhibits two kinks at $\Lambda = 10.2$ and $15.0$. 
}
\label{fig:struct}
\end{figure*} 

\subsection{Shear modulus}

Within linear response, the shear modulus is given by $\mathcal{G} = \sigma^{tot}_{xy}/{\cal E}_{xy}$. We compute this from numerical simulations by applying a shear strain ${\cal E}_{xy}$ and measuring the change in total shear stress $\sigma_{xy}^{\mathrm{tot}}$ accounting for trajectories displaying linear response; see Appendix-\ref{sec_shear}.
The stress tensor has contributions from interaction calculated using the usual virial term and swim stress due to activity.  
The standard virial expression due to interaction gives, 
\begin{equation}
    \sigma^{vir}_{\alpha \beta} = \frac{1}{d V} \sum_{i=1}^{N} \sum_{j=1}^{neb(i)} f_{ij,\alpha} \, r_{ij,\beta}, 
\end{equation}
where, $V$ denotes the system volume in $d=2$ dimensions, $f_{ij}$ is the interaction force between particles $i$ and $j$ with separation $r_{ij}$, and $\alpha$, $\beta$ denote the Cartesian components of the vectors. On the other hand, the summation index $j$ runs over all interacting neighbors of the $i^\mathrm{th}$ particle.
The active swim stress  
can be expressed as~\cite{Solon2015f}, 
\bea 
\sigma^{sw}_{\a \be} = \frac{\g v_0}{D_r V} \sum_{i=1}^N \big\la \dot r_{i, \a} \, n_{i, \be} \big\ra . 
\eea
The total shear stress is then calculated as $\sigma^{tot}_{xy} = \sigma^{vir}_{xy}+\sigma^{sw}_{xy}$. Appendix-\ref{sec_shear}
outlines details of the shear modulus calculation. 
{The mean swim stress does not change under shear, as confirmed by numerical calculations; so the shear modulus is set solely by the virial stress.}

Figure~\ref{fig:struct}(a) shows the change in mean shear modulus, $\cal G$, with standard errors, as a function of activity $\Lambda$. 
We observe a steady decrease in it from the equilibrium value, indicating softening of the solid with activity. As activity increases, the fraction of trajectories following linear response decreases to vanish near $\L=10$; see Appendix-\ref{sec_shear}.  
The decrease in $G$ with activity can be described by 
${\cal G} = {\cal G}_0 - {\cal G}_1 \Lambda^2$ with ${\cal G}_0 = 109.84$ and ${\cal G}_1 = 0.094$.
To further explore this softening and the fluidization transition, we compute structural indicators, including solid order $\psi_{\mathbf{G}}$, hexatic order $\psi_6$, and defect fraction $\delta_d$.

\begin{figure}[t!] 
\centering
\includegraphics[width=0.9\linewidth]{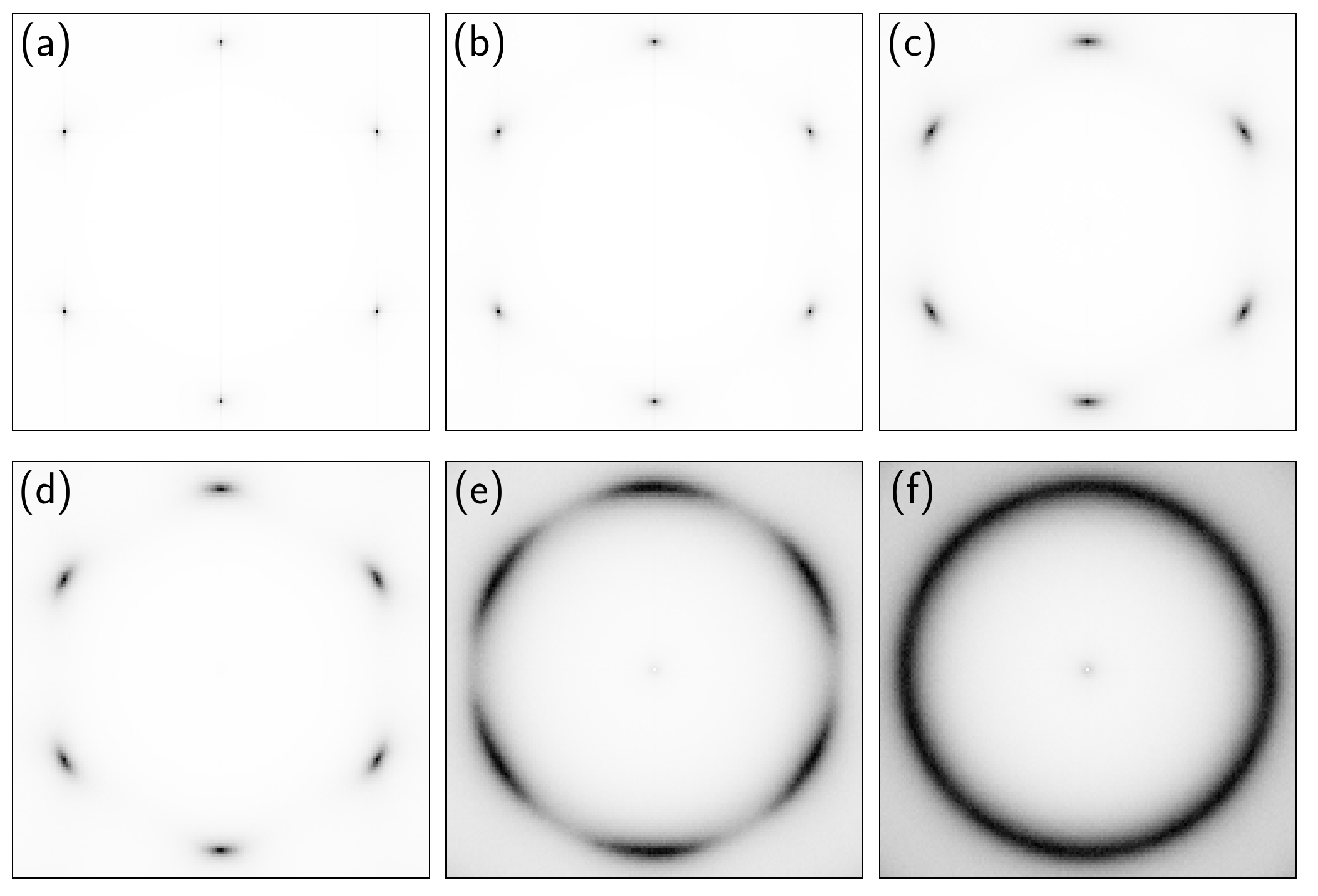}   
\caption{The structure factor $S(\mathbf{q})$ at $\mathrm{\Lambda}=4$~(a), $10$~(b), $10.2$~(c), $10.5$~(d), $15$~(e), and $20$~(f). }
\label{fig:Sq}
\end{figure}

\subsection{Solid order}
The structure factor of the system is given by
\bea
S({\bf q}) = \la \r_{\bf q} \r_{-{\bf q}} \ra
\eea
where $\r_{\bf q} = \f{1}{N} \sum_{i=1}^N \exp( i {\bf q} \cdot \rv_i)$. Plots of $S({\bf q})$ at representative $\L$ values are shown in Fig.\ref{fig:Sq}. 
In a perfect triangular lattice, quasi-Bragg peaks emerge at $ {\bf q}_p = (0, \pm 2\pi/a_y), (\pm 2\pi/a, \pm \pi/a_y) $, as shown in Fig. \ref{fig:Sq}(a),(b). 
{ The solid order parameter $ \langle \psi_{\bf G} \rangle $ is calculated by taking an average over $ S({\bf q}) $ values calculated at the six quasi-Bragg peaks located at $ {\bf G}:= \{ {\bf q}_p \} $}:
\bea
\langle \psi_{\bf G} \rangle = \f{1}{6}\sum_{p=1}^6 \left\la  \left | \f{1}{N} \sum_{i=1}^N \exp( i {\bf q}_p \cdot \rv_i) \right |^2 \right\ra
\eea

 Fig. \ref{fig:struct}(b) shows the variation of $ \langle \psi_{\bf G} \rangle $ with $ \Lambda $. The order parameter $ \langle \psi_{\bf G} \rangle $ vanishes as the solid melts. At the melting point
$ \Lambda = 10.0 $ the variance of the order parameter $ \langle (\Delta \psi_{\bf G})^2 \rangle $ peaks (inset of Fig.~\ref{fig:struct}(b)\,). This analysis shows that the quadratic scaling of non-affinity with activity is observed across the entire activity range before solid melting.

As the active solid softens and melts with increasing $ \Lambda $, the quasi-Bragg peaks flatten and merge (Fig.~\ref{fig:Sq}(c)-(f)\,), similar to equilibrium 2D melting~\cite{Khali2020}. The $S({\bf q})$ in Fig.~\ref{fig:Sq}(c),(d) are characteristic of a hexatic. The sixfold symmetry softens and starts to merge in Fig.~\ref{fig:struct}(e) as the hexatic melts. Finally, at $\L=20$, $S({\bf q})$ gets the uniform ring reflecting isotropy of the fluid; see  Fig.~\ref{fig:Sq}(f). Therefore, $S({\bf q})$ suggests a two-stage melting from solid to hexatic to fluid.

\subsection{Hexatic order}

This section examines the change in the hexatic order parameter to distinguish hexatic melting from solid melting. Each particle in the system can be associated with a hexatic bond orientational order. 
For the $k$-th particle, it is defined as $\psi_6^k = (1/n) \sum_{i=1}^n e^{i 6 \h_{kj}}$, where $\h_{kj}$ is the bond angle between particle $k$ and its topological (Voronoi) neighbor $j$  relative to the $x$-axis, and $n$ is the number of such neighbors.
In Fig.~\ref{fig:struct}(c), we plot the mean squared amplitude of the hexatic order parameter of the system~\cite{Khali2021}
\bea
    \langle\psi_6\rangle = \left\langle \left| \frac{1}{N} \sum_{k=1}^{N} \psi_6^k \right|^2 \right\rangle, \eea
as a function of $\Lambda$. The hexatic order diminishes with $\Lambda$, but remains significant beyond  $\Lambda = 10$, suggesting the system stays in the hexatic phase past the solid melting point, with the order vanishing near $\L=20$.  The hexatic melting point $\L=15$ is identified by the peak in the variance of hexatic order $\langle (\Delta \psi_6)^2 \rangle$ (inset of Fig.~\ref{fig:struct}(c)\,).

\begin{figure}[t!] 
\centering
\includegraphics[width=\linewidth]{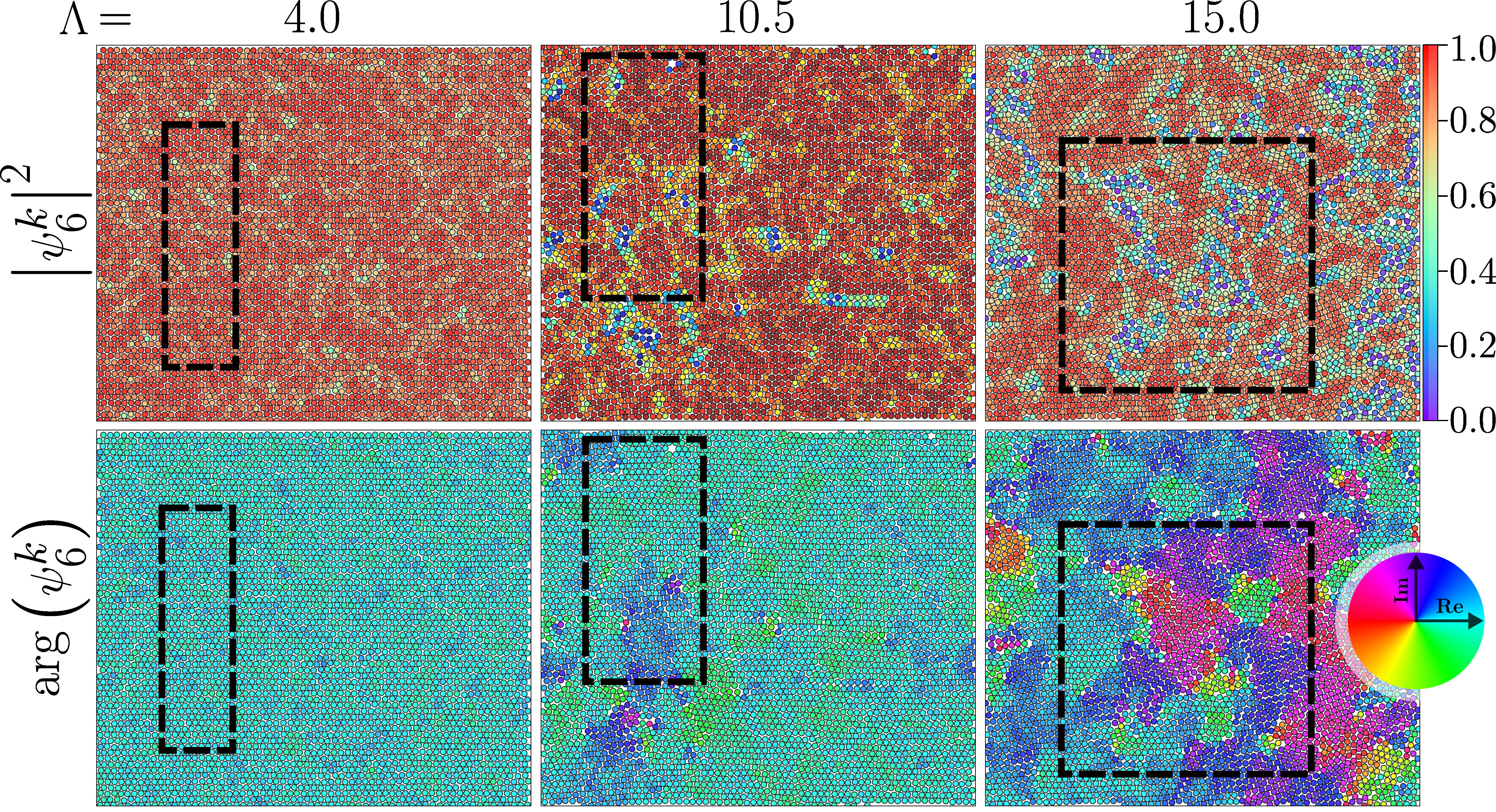} 
\caption{%
   Typical configurations with local hexatic order at $\Lambda = 4$, $10.5$, and $15$: top row shows $|\psi^k_6|^2$ and bottom row shows $\mathrm{arg}(\psi^k_6)$. Fig.~\ref{fig:voronoi} displays topological defects in the selected regions denoted by boxes in the figures.
    }
\label{fig:fullconf}
\end{figure} 

The local squared amplitudes $|\psi_6^k|^2$ and arguments arg($\psi_6^k$) of hexatic order for typical configurations across the transition are shown in Fig.~\ref{fig:fullconf} as the active solid softens and melts with increasing $\Lambda$. The drops in local hexatic amplitude, linked to hexatic orientation shifts, signal topological defect formation, as detailed in the following section.

\begin{figure}[ht!] 
\centering
\includegraphics[width=8.6cm]{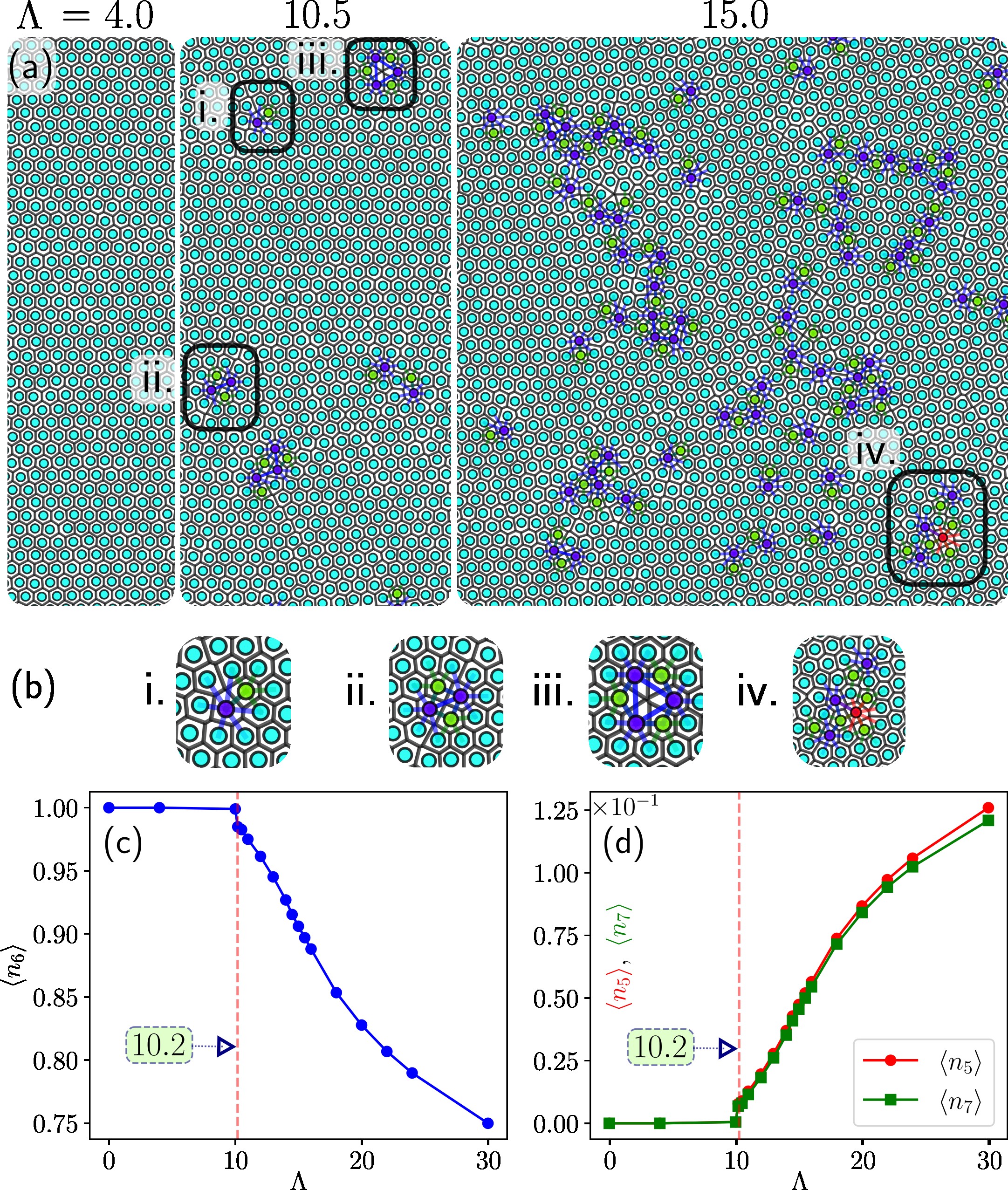} 
\caption{%
{(a)} Voronoi tessellations for selected regions (see Fig.~\ref{fig:fullconf}) at $\mathrm{\Lambda}=4.0$, $10.5$, and $15.0$. Particles are colored by their coordination number: green for five-fold, blue for seven-fold, and red for eight-fold, while sixfold-coordinated particles remain uncolored for clarity.
    {(b)} Regions containing defects are highlighted:
     {(i)} an isolated 5--7 defect;
     {(ii)} a pair of adjacent 5--7 defects;
     {(iii)} a cluster of three 5--7 defects surrounding a void; and
     {(iv)} a more complex defect cluster.
    {(c)} The fraction of sixfold-coordinated particles is plotted as a function of $\mathrm{\Lambda}$. 
    {(d)} The fractions of particles with five- and seven-fold coordination are shown as functions of $\mathrm{\Lambda}$.
 }
\label{fig:voronoi}
\end{figure}

\subsection{Topological defects}

We use Voronoi tessellations to identify each particle's topological neighbors~\cite{freud2020}. In a perfect lattice, each particle has 6 neighbors, with 5- or 7-fold deviations indicating topological defects.
Minimal fluctuations in solid accommodate bound 5–7–5–7 dislocation pairs. According to equilibrium BKTHNY theory, solid-hexatic melting occurs via the unbinding of these dislocation pairs, creating free dislocations, while the transition from hexatic to liquid is driven by the unbinding of dislocations into free 5- or 7-fold disclinations.  
{
Since a detailed characterization of BKTHNY-like melting in active solids has been presented previously~\cite{Shi2023, Paliwal2020}, we do not revisit measurements of order-parameter correlations or scaling exponents here.
}

In Fig.~\ref{fig:voronoi}(a),(b), we identify such defects in typical configurations across activated solid melting. We identify clusters of such defects, including five, seven, and eight-fold defects. We use the fraction of particles with $\nu$ neighbors, $n_\nu = \langle N_\nu \rangle/N$ to analyze the defect formation. Here $N_\nu$ indicates the number of particles with $\nu=5,6,7$ neighbors and $\la \dots \ra$ denotes the steady state average. Although larger departures from $\nu=6$ are present, their fraction is significantly low. 

As activity $\Lambda$ exceeds $\L=10.0$, the solid melting point, $\langle n_6 \rangle$ decreases, while the defect fraction, $\langle n_{5,7} \rangle$, increases (see Fig.~\ref{fig:voronoi}(c),(d)). The mean fraction of all particles with non-six neighbors, $\la \delta_d \ra= \frac{1}{N} \sum_{\nu \neq 6} \la N_\nu \ra$, becomes non-zero at the solid melting point and increases with higher activity (Fig.~\ref{fig:struct}(d)). The fluctuation in $\delta_d$ peaks twice: first at $\L=10.2$, near the solid melting point, and again at $\L=15.0$, at the hexatic melting point, reflecting enhanced defect fluctuations at both the melting transitions (see inset of Fig.~\ref{fig:struct}(d)).

\section{Local softening: control of non-affinity and defect}
\label{control}

Controlling local mechanical properties in solids is a longstanding challenge in physics and materials science~\cite{baluffi2012}, particularly for colloidal crystals where structure and elasticity can be tuned at the particle scale~\cite{ivlev2012, spalding2008}. 
{
Earlier attempts to control non-affine fluctuations in passive solids proposed introducing a force field conjugate to non-affine modes as a theoretical route to control local non-affinity~\cite{ganguly2015statistics, popli2018translationally}. This field corresponds to an effective many-body force that depends explicitly on the instantaneous particle configuration within a local neighborhood, with fixed reference lattice positions entering as parameters. The implementation scheme based on dynamically modulated laser traps outlined in~\cite{ganguly2015statistics}, therefore, requires real-time particle tracking and the synthesis of highly structured, configuration-dependent forces, making experimental realization extremely challenging.

In the present work, building on our analysis of activity-induced non-affine fluctuations, we propose a qualitatively distinct and experimentally accessible strategy for controlling non-affinity through locally tunable activity. Specifically, optically activated self-thermophoretic Janus colloids~\cite{Jiang2010}, combined with a simple defocused laser beam, can be used to locally activate a region of an otherwise passive colloidal crystal, thereby robustly inducing non-affine deformations. Similar localized activation protocols are feasible in a broad class of active colloidal and robotic systems~\cite{bechinger2016active, Yu2021}. This approach enables direct spatial modulation of the activity strength $\Lambda$, which in turn locally controls the degree of non-affinity according to $\langle X \rangle \sim \Lambda^{2}$.  Crucially, this approach circumvents the need for real-time configuration tracking or the synthesis of complex many-body forces, and instead relies on the system's intrinsic active dynamics to achieve local, tunable control of non-affine fluctuations.
}

\begin{figure}[t!] 
\centering
\includegraphics[width=8.6cm]{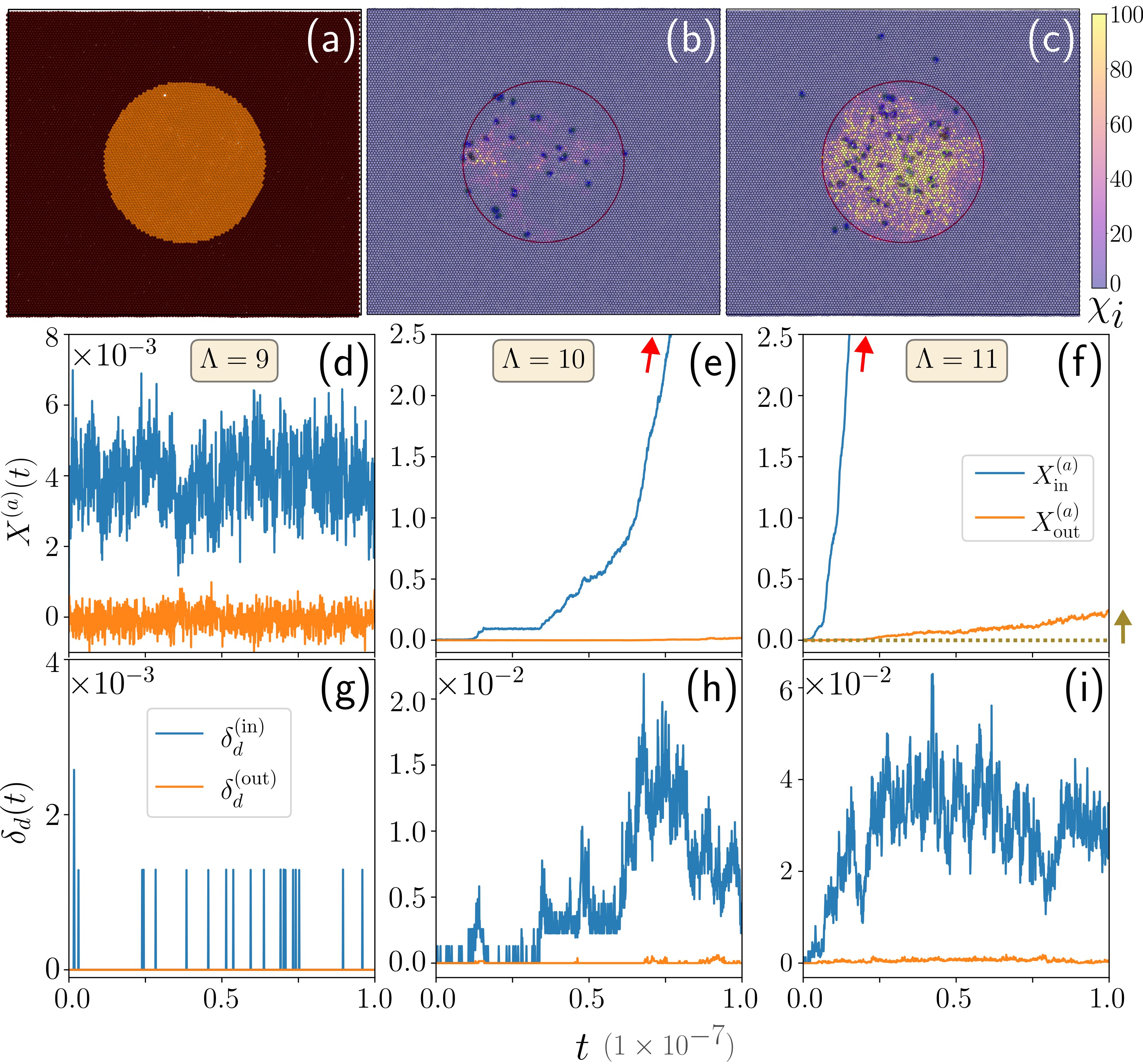}
\caption{%
(a) Schematic of the simulation setup, where particles are selectively activated within a circular region of radius $r_{\rm beam} = 30\sigma$ at the center of the simulation box, simulating laser irradiation.  
(b), (c) Snapshots of the system under $\L=11$ at times $t = 2 \times 10^6 \tau_p$ and $6.02 \times 10^6 \tau_p$, with particles colored by their local non-affinity $\chi_i$ and defects (particles with non-six neighbors) indicated.  
(d)–(f) Time evolution of the excess non-affinity $X^{(a)}$ inside (blue) and outside (orange) the active region.  
(g)–(i) Evolution of defect fraction inside (blue) and outside (orange) the active region. Parameter values are $\Lambda = 9$ (d,g), $\Lambda = 10$ (e,h), and $\Lambda = 11$ (f,i).
}
\label{fig:laser}
\end{figure}

To investigate these effects near the solid–hexatic transition, we simulate a large system of $N = 16384$ particles arranged initially in a triangular lattice at density $\rho \s^2 = 1.1$, coupled to a thermal heat bath with diffusivity $\tilde{D}_t = 1$, and a fixed $\tilde D_r=3.0$. The system evolves in a rectangular 
simulation box of size $131.14\sigma \times 113.57\sigma$, under periodic boundary conditions. After sufficient equilibration, activity is applied selectively within a central circular region (radius $r_{\rm beam} = 30\sigma$), as illustrated in Fig.~\ref{fig:laser}(a). We explore three activity levels: below ($\Lambda = 9$), at ($\Lambda = 10$), and above ($\Lambda = 11$) the transition threshold. 
The activity of each particle $\Lambda_i(t)$ is implemented via a binary field $\s_i(t)$, such that:
\[
\Lambda_i(t) = \sigma_i(t)\Lambda,
\]
where $\s_i(t)=1$ if particle $i$ is located within the active region at time $t$, and $\sigma_i (t) = 0$ otherwise. This formulation ensures that activity is always applied in the spatially defined region, regardless of the identity of individual particles. The orientation of the active force undergoes rotational diffusion with a fixed rate $\tilde D_r=3.0$, consistent with the standard active Brownian particle model (Eq.\eqref{eq_Lange}). System properties such as the non-affine parameter $\chi_i$,  global non-affinity $X$, and defect fraction are computed at regular intervals.

A schematic of this setup is shown in Fig.~\ref{fig:laser}(a). 
Snapshots for $\Lambda = 11$ in Fig.~\ref{fig:laser}(b,c) show evolution over time, where particles are colored by their local non-affinity $\chi_i$, display pronounced softening and defect formation confined within the activated zone, while $\chi_i$ remains near equilibrium outside. At early times, defects remain localized in the active zone, but later, isolated defects appear in the passive region due to migration and delayed annihilation of those originating from the active zone. To quantify structural changes, we define $X^{(a)} = X(\Lambda) - X(0)$, the shift in global non-affinity from equilibrium, as defined in Sec.~\ref{subsec:scaling_X_sim}. 
The evolution of this quantity in the active zone $X^{(a)}_{\rm in}$ and in the passive zone $X^{(a)}_{\rm out}$ is shown in Fig.~\ref{fig:laser}(d)--(f).
These figures show that below the transition, $X^{(a)}_{\rm in}$ fluctuates around a steady mean with enhanced amplitude, while $X^{(a)}_{\rm out}$ stays negligible. At the transition, $X^{(a)}_{\rm in}$ diverges, with no corresponding rise in $X^{(a)}_{\rm out}$, indicating localized melting. Above the transition, divergence in $X^{(a)}_{\rm in}$ occurs earlier and is followed by an increase in $X^{(a)}_{\rm out}$, suggesting outward propagation of non-affineness.

The evolution of defect fraction in Fig.~\ref{fig:laser} (g-i) complements these observations. For $\Lambda = 9$, defects intermittently form and annihilate; for $\Lambda \geq 10$, the defect fraction increases significantly within the activated region and more gradually outside it.  
{The simultaneous rise of non-affinity and defect fraction suggests a coupled onset of mechanical disorder at these high activities.}


\section{Discussion and Conclusion}
\label{conc}

In this work, we have investigated spontaneous deformations in active solids by quantifying local rearrangements through non-affine fluctuations relative to global strain. 
{
The observed scalings of the global non-affinity with activity parameters reflect the distinct mechanisms driving particle rearrangements in active solids.
Increasing the self-propulsion speed enhances active fluctuations, thereby increasing non-affinity, while increasing persistence allows particles to maintain their heading over longer times, promoting more coherent local rearrangements. At low to moderate persistence, this leads to a rise in the mean global non-affinity, but at high persistence, the system becomes jammed, so further increases in persistence no longer enhance non-affinity, reflecting the interplay between persistent driving and structural constraints. These mechanisms differ fundamentally from equilibrium solids, where particle displacements undergo Gaussian fluctuations, resulting in chi-squared distributions of local non-affinity, with correlations decaying within roughly two lattice spacings. 
In contrast, the persistent local forces in active solids produce non-Gaussian fluctuations, longer-ranged non-affine correlations, and distributions of non-affinity that deviate markedly from the chi-squared form, revealing cooperative particle rearrangements absent in equilibrium. 
}

%
{
Our main findings can be summarized in three key points. First, combining scaling arguments and numerical simulations, we quantified the steady-state mean non-affinity as a function of activity parameters. It grows quadratically with active speed and increases linearly with persistence before saturating, a behavior not captured by simple active-temperature estimates.
Second, at higher activity, local non-affinity distribution broadens, becomes more skewed with heavy tails, and develops a secondary peak, reflecting the coexistence of rare large rearrangements and predominantly small fluctuations. This coexistence precedes the divergence of non-affinity, serving as a precursor to defect proliferation and eventual melting. 
Spatial correlations of non-affinity exhibit a correlation length that grows with active speed and increases with persistence before saturating, while the dynamical relaxation of the global non-affinity is set by the persistence time. Activity-induced softening reduces the shear modulus and destabilizes both solid and hexatic order, with two-step melting mediated by proliferation of topological defects.  Non-affine fluctuations provide a sensitive probe of local rearrangements and defect precursors in crystalline solids, but their presence alone does not imply melting; rather, they signal regions where defect nucleation and subsequent structural transformations may occur.

Third, we demonstrated that spatially patterned activation provides a direct and experimentally feasible route to locally induce non-affinity and mechanical softening. By selectively activating regions of a passive colloidal crystal -- for example, via optically controlled self-thermophoretic Janus particles -- non-affine deformations can be reliably induced without complex many-body interactions or real-time particle tracking. This approach allows tunable, localized control of mechanical response and opens the way for spatially programmable solid mechanics. 
}  

Beyond deepening our understanding of active solid behavior, these results have broad implications for the design of adaptive metamaterials and may inform how biological systems dynamically regulate their rigidity through internally generated stresses.

\section*{Author Contributions} 
The work began with a discussion between DC and SS, followed by initial non-affinity scaling simulations carried out by DD. PN set up and conducted all the final numerical simulations and analyzed the results. DC supervised the project and wrote the final manuscript; all authors contributed to initial writing and discussed the results.

\section*{Data availability }
The data are available from the authors upon reasonable request.

\section*{Conflicts of interest}
There are no conflicts of interest to declare. 

\acknowledgments
DC acknowledges support from the Department of Atomic Energy (OM no. 1603/2/2020/IoP/R\&D-II/15028), an Associateship at ICTS-TIFR, Bangalore, and thanks MPIPKS, Dresden, for hospitality during a two-month visit in 2024, where part of this work was done. 
DC also thanks Tamoghna Das for useful comments. 
Numerical simulations were performed using the SAMKHYA high-performance computing cluster and other computational resources at the Institute of Physics, Bhubaneswar.

This work is dedicated to the memory of our late co-author, Prof. Surajit Sengupta~\cite{surajitweb}, in recognition of his lasting contributions to the field and to this research.

\appendix

\begin{figure}[t!]
    \centering
    \includegraphics[width=7cm]{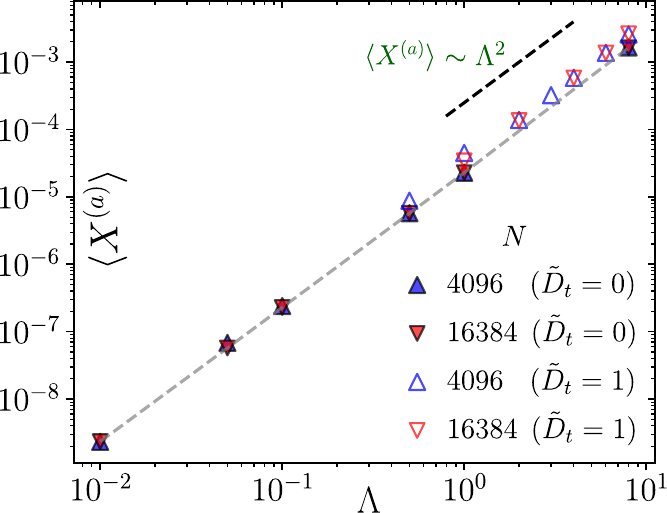}
    \caption{ As in Fig.~\ref{fig:scaling_X}, we plot $\langle X^{(a)} \rangle$ for two system sizes ($N=4096, 16384$) at density $\rho\sigma^2 = 1.1$, demonstrating that the scaling $\langle X^{(a)} \rangle \sim \Lambda^2$ holds for both the thermal ($\tilde{D}_t=1$) and athermal ($\tilde{D}_t=0$) cases at the larger system size. }
    \label{fig:Xscale_multi_N}
\end{figure}

\section{Dimensionless equations}
\label{app_dim}
To perform numerical simulations, we use the dimensionless equations by applying the following substitutions:
 $\rv_i \to \tilde \rv_i = \rv_i/\s$, 
 $t \to \tilde t=t/\t_u = t \mu \e/\s^2$, 
 the dimensionless activity $\L= v_0\t_u/\s = v_0 \s/\mu \e$, 
 the dimensionless translational diffusivity $D_t \to \tilde D_t= D_t \t_u/\s^2 = D_t/\mu \e$, 
 $\nabla_i \to \tilde \nabla_i = \s \nabla_i$,  
 $U(r_{ij}) \to \tilde U= U/\e$, 
 $\tilde \mu = \mu \e \t_u/\s^2= 1$ and 
 the dimensionless rotational diffusivity $\tilde D_r= D_r \s^2/\mu \e$. 
 By using these substitutions, one obtains the dimensionless form of the equations
\begin{align}
    d \tilde \rv_i(\tilde t) &= \L \, \hat{\bf n}_i(\tilde t) d \tilde t -  \tilde \nabla_i\sum_{j \epsilon R_i} \tilde U(r_{ij}) d \tilde t + \sqrt{2 \tilde D_t}\, d{\bf B}_i(\tilde t) \nn \\
    d{\theta_i}( \tilde t) &=  (2 \, \tilde D_r)^{1/2}\,dB^{r}_i( \tilde t). \nn
\end{align} 

We perform numerical simulations with $\tilde D_t=1$ for thermal, and $\tilde D_t=0$ for athermal cases.

\section{Scaling ansatz for $X$ at different system sizes}
\label{app_scaling}

In Fig.~\ref{fig:scaling_X} of the main text, we showed the scaling behavior $\langle X^{(a)} \rangle \sim \Lambda^2$ for a system of size $N=4096$. To verify the robustness of this scaling, we present additional data for a larger system size ($N=16384$) at density $\rho\sigma^2 = 1.1$ in Fig.~\ref{fig:Xscale_multi_N}. We confirm that the quadratic scaling $\langle X^{(a)} \rangle \sim \Lambda^2$ remains valid for both thermal ($\tilde{D}_t = 1$) and athermal ($\tilde{D}_t = 0$) cases, consistently across system sizes.

\begin{figure}[t!] 
\centering 
\includegraphics[width=1.0\linewidth]{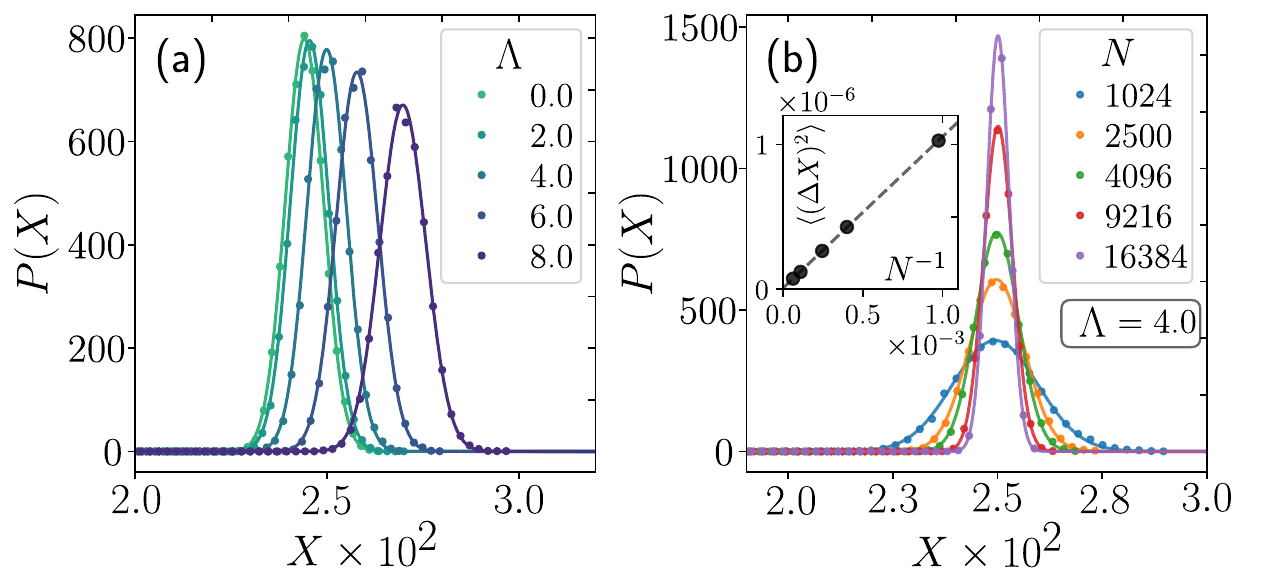}
.\caption{ (a)~Probability distribution $P(X)$ for $\rho=1.1$ at different $\Lambda$ values. Both mode and variance increase with activity. (b)~$P(X)$ for different system sizes at $\Lambda=4$. Variance $\s^2$ of $P(X)$ showing $\s^2 \sim N^{-1}$. } 
\label{fig:Prob_X}
\end{figure}  

\section{Probability distributions of $X$}
\label{sec_Px}

In Figure~\ref{fig:Prob_X}, we explore the probability distributions of the global non-affine parameter $X$ for an ABP solid at $\tilde D_r=3.0$, $\tilde{D_t} = 1$ and $\rho \s^2 = 1.1$. Figure~\ref{fig:Prob_X}(a) compares $P(X)$ for $N=4096$ particles across different $\Lambda$ values, showing that as $\Lambda$ increases, the mode of the distribution shifts to larger $X$ (consistent with the $\la X \ra \sim \L^2$ scaling) and broadens {as $\la X^2 \ra - \la X \ra^2 \sim \L^4$ (plot not shown)}. This behavior captures softening even in the absence of defect formation (as noted in the main text, the onset of defect formation appears at $\L\approx 10$). 
As shown in Fig.~\ref{fig:Prob_X}(b) for $\Lambda=4$ and $N$ ranging from 1024 to 16384, the probability distribution $P(X)$ is nearly Gaussian with variance $\sigma^2 \sim 1/N$, consistent with the central limit theorem for the mean of uncorrelated entities. {The vanishing variance of  $X$ with $N$ characterizes the intensiveness of the global non-affine parameter, characterizing the particle rearrangements expelling local stress.}  Similar scaling behavior for fluctuations in $X$ was observed before in equilibrium solids~\cite{ganguly2015statistics}. 

\begin{figure}[t]
    \centering
    \includegraphics[width=0.95\linewidth]{{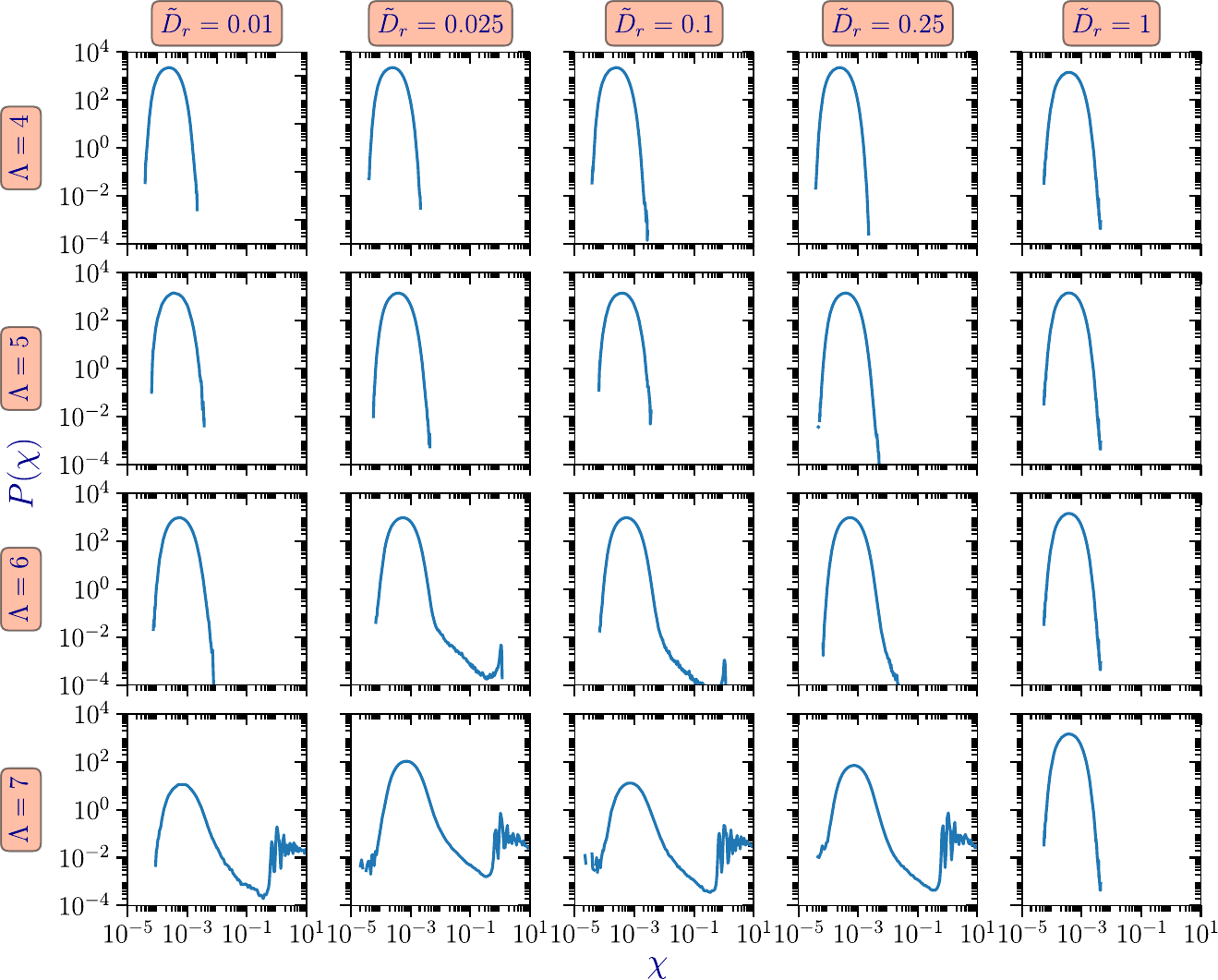}}
    \caption{
    {
    Stacked probability distributions $P(\chi)$ of local non-affinity on log-log axes for varying activity $\Lambda$ (rows) and rotational diffusion $\tilde D_r$ (columns). Increasing $\Lambda$ shifts and broadens the distributions, while increasing $\tilde D_r$ suppresses large-$\chi$ tails; bimodality emerges at large $\Lambda$ for $\tilde D_r<0.25$.
    }
    }
    \label{fig:Pchi_app}
\end{figure}

{
\section{Probability distributions of local non-affinity}
\label{app_Pchi}

Figure~\ref{fig:Pchi_app} compiles stacked probability distributions $P(\chi)$ of the local non-affinity $\chi$ on log-log axes over a broad range of active speed $\Lambda$ and rotational diffusivity of the heading direction $\tilde D_r$, for an athermal system ($\tilde D_t=0$). Rows correspond to increasing activity, $\Lambda=1,4,5,6,7$, while columns span rotational diffusivities $\tilde D_r=0.010,0.025,0.100,0.250,1,3,10,30,100,500$. At fixed active speed (across each row), increasing $\tilde D_r$ has a nuanced impact on the distribution of non-affinity. At high enough active speed, e.g., $\L=6$, extremely large persistence ($\tilde D_r=0.01$) keeps the distribution unimodal, yet produces a tail towards higher $\chi$ over an intermediate range of persistence. While persistence helps produce bimodality, too large a persistence can lead to jamming, suppressing the large $\chi$ fluctuations.  Beyond this regime, a progressive decrease in persistence suppresses the large-$\chi$ tail, and the distributions become unimodal at high $\tilde D_r$.

Conversely, at fixed $\tilde D_r$ (down each column), increasing $\Lambda$ shifts probability weight toward larger $\chi$ and broadens $P(\chi)$. For low rotational diffusivity, $\tilde D_r<0.25$, this trend culminates in a clear bimodality at large $\Lambda$, reflecting the coexistence of a dominant low-$\chi$ population with a secondary high-$\chi$ contribution arising from rare, large non-affine rearrangements. Near $\tilde D_r=0.25$, the distributions show a strong enhancement of typical $\chi$, marked by a rightward shift and broadening of the main peak, signaling a crossover to the unimodal, tail-suppressed regime observed at larger $\tilde D_r$.
}

\begin{figure}[t!] 
\centering
\includegraphics[width=8cm]{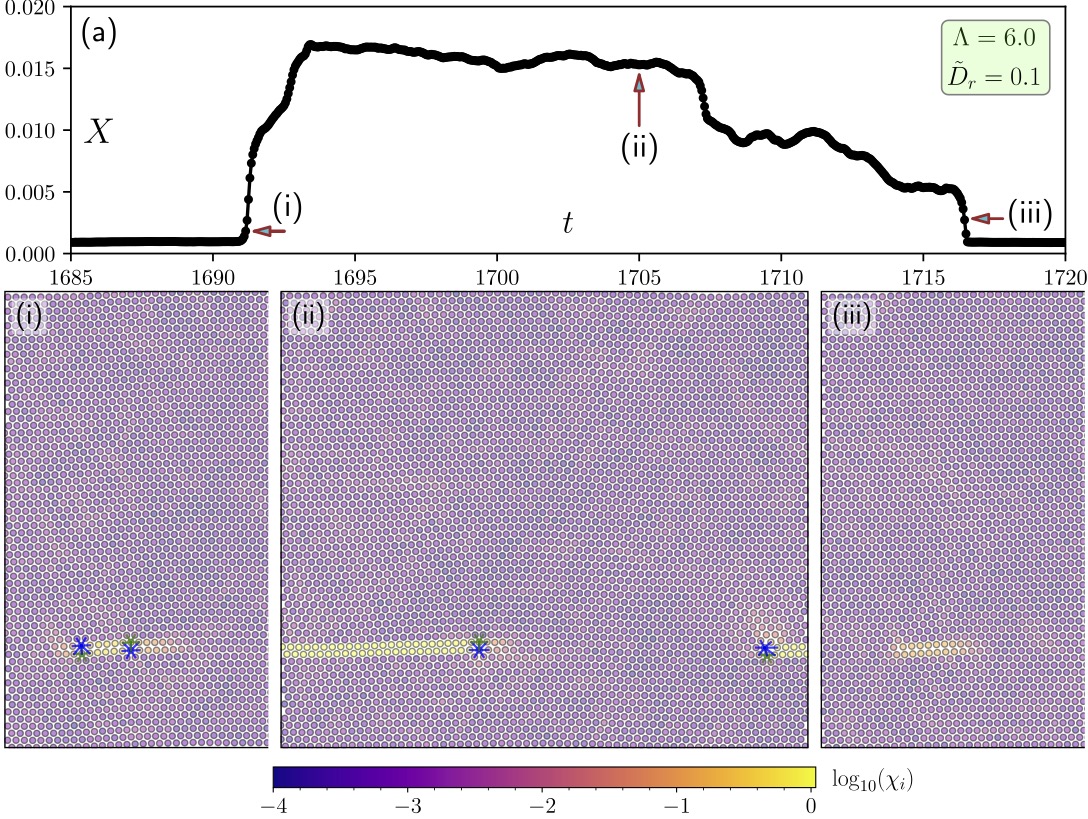} 
\caption{
(a) Representative time series of the global non-affinity $X$ at $\Lambda=6.0$ and $\tilde D_r=0.1$, showing the spontaneous emergence and decay of highly non-affine regions.
(b) Configurations at the marked times (i)–(iii) in (a). Particles are colored by $\log_{10}(\chi_i)$, with defects indicated as in Fig.~\ref{fig:laser}(b,c). Panels (i) and (iii) show cropped (half-width) views around the event, while (ii) shows the full simulation box. 
    }
\label{fig:conf_coex}
\end{figure} 

\begin{figure}[tbp!] 
\centering
\includegraphics[width=8cm]{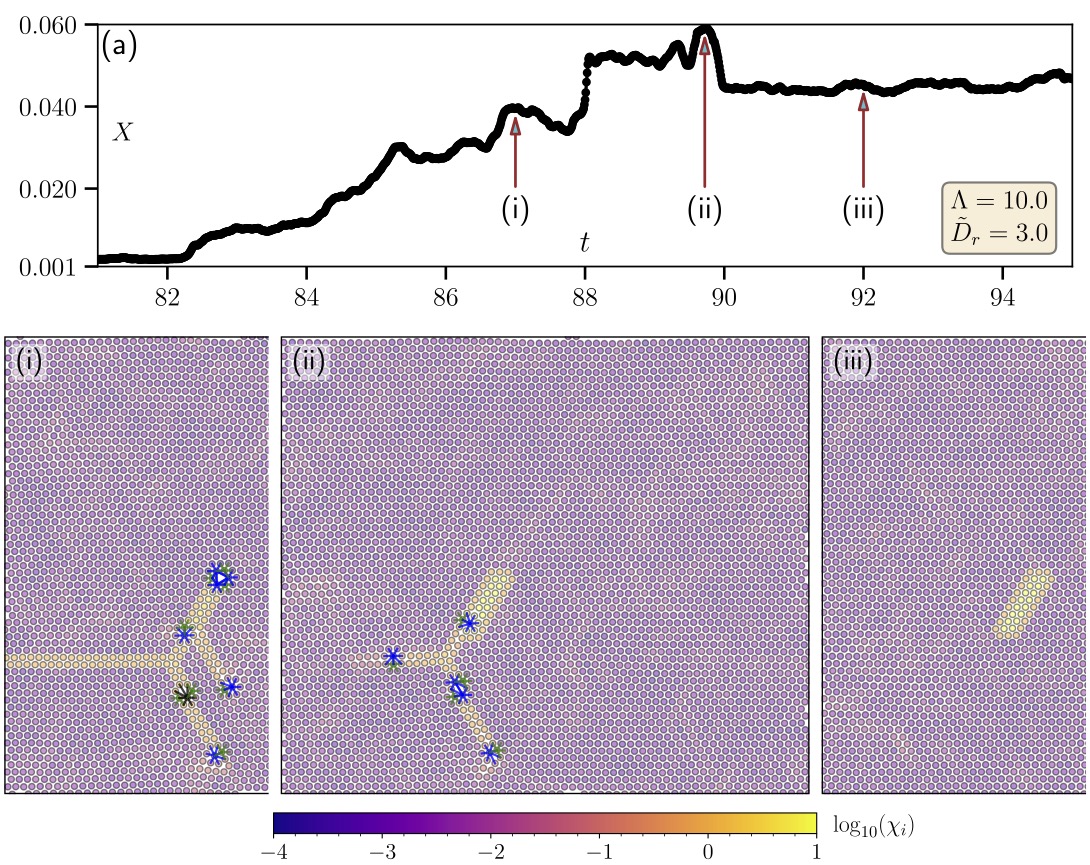}   
\caption{
    (a) Similar to Fig.~\ref{fig:conf_coex}, the time series of the global non-affinity $X$ at $\Lambda=10.0$ and $\tilde D_r=3.0$, configurations at the three marked times (i)--(iii) in the time series $X(t)$ shown at $t=87\,\tau_u$, $89.72\,\tau_u$, and $92\,\tau_u$, respectively. Particles are colored by $\log_{10}(\chi_i)$, with defects indicated as in Fig.~\ref{fig:laser}(b,c). 
    }
\label{fig:conf_coex2}
\end{figure}

\section{Non-affine configurations}
\label{sec_conf_coex}
{
To better understand the bimodal distributions $P(\chi)$ in Fig.~\ref{fig:Pchi} and Fig.~\ref{fig:Pchi_app} for athermal systems ($\tilde D_t = 0$), we show representative time series of $X(t)$ together with illustrative configurations for two parameter sets. The bimodality in $P(\chi)$ arises from both intermittent plastic bursts and steady-state coexistence of non-affine zones within an otherwise affine background. Fig.~\ref{fig:conf_coex}(a) shows a burst event at $\Lambda=6.0$ and $\tilde D_r=0.1$. The corresponding particle configurations in Fig.~\ref{fig:conf_coex}(b), colored by local non-affinity, resolve the sequence: nucleation of a localized high-$\chi$ region, growth into an extended slip-like band, and partial relaxation toward the low-$\chi$ background (Fig.~\ref{fig:Pchi}(b)). In contrast, Fig.~\ref{fig:conf_coex2} illustrates the formation and growth of branched non-affine zones that eventually stabilize, producing steady-state coexistence of local non-affine rearrangements with an otherwise affine solid. These examples are illustrative: both transient bursts that fully relax and events that stabilize into persistent non-affine zones contribute to the bimodal distributions observed across the parameter regimes.
}

\section{Obtaining shear modulus}
\label{sec_shear}

For each $\Lambda$, we initialize with a triangular lattice and run simulations for $t_{\max} = 10^{7}$ with a time step $dt = 10^{-5} \t_p$ until steady state is reached. We then apply three deformation stages:
$${\cal E}_{xy} \Rightarrow 0 \;\rightarrow\; +{\cal E}_{xy}^{\max} \; \rightarrow\; -{\cal E}_{xy}^{\max} \; \rightarrow\; 0,$$
at a strain rate $\dot{{\cal E}} = 10^{-2} D_r$ and ${\cal E}_{xy}^{\max} = 0.035$. The shear modulus is a linear response coefficient; thus the chosen ${\cal E}_{xy}^{\max}$ is to ensure the system response stays within the linear regime for accurate shear modulus extraction.

{We perform the calculations at $\tilde D_t=1.0$ and $\tilde D_r=3.0$.}
For $\Lambda \leq 6$, all four initial configurations exhibit linear stress-strain behavior. For $\Lambda > 6$, non-linear deviations appear. We generate $n_{\mathrm{tot}} \approx 100$ initial configurations, discarding those with nonlinearity for shear modulus analysis, which becomes more common at higher $\Lambda$. The shear modulus is then computed by averaging the linear response data from the remaining $n_{\mathrm{lin}}$ hysteresis-free ensembles, as shown in Fig.~\ref{fig:shear_ens}(a). 
{It displays the total shear stress, $\sigma_{xy}^{\mathrm{tot}}$, as a function of shear strain, $\mathcal{E}_{xy}=\gamma$, at $\Lambda=7$ for 24 independent ensembles. All trajectories exhibit a linear response without noticeable hysteresis. Different colors denote individual ensembles, while the black line is a linear fit whose slope yields the shear modulus, $\langle {\cal G} \rangle = 105.7 \pm 0.679$ (in units of $\epsilon/\sigma^2$). Here we reiterate that the mean swim stress remains unchanged under shear, as confirmed by numerical calculations; thus, the shear modulus is determined solely by the virial stress.

\begin{figure}[tbp!] 
\centering
\includegraphics[width=8.2cm]{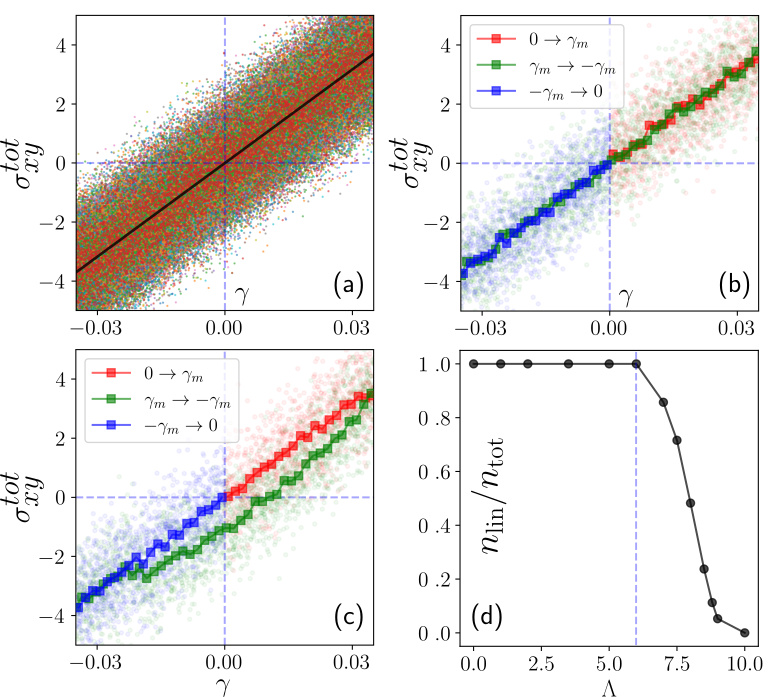}   
\caption{
{
Shear-stress response under cyclic shear. (a)~Total shear stress $\sigma_{xy}^{\mathrm{tot}}$ vs shear strain $\gamma$ at $\Lambda=7$ for 24 ensembles; the linear fit (black) yields $\langle {\cal G} \rangle = 105.7 \pm 0.679$ ($\epsilon/\sigma^2$). (b,c)~Representative linear and nonlinear (hysteretic) trajectories; nonlinear cases are excluded from the modulus estimate. Symbols denote raw data (semi-transparent circles) and averages over 100 points (squares); colors indicate shear direction. (d)~Fraction of linear ensembles, $n_{\mathrm{lin}}/n_{\mathrm{tot}}$, vs activity $\Lambda$.
}
}
\label{fig:shear_ens}
\end{figure} 

Panels (b) and (c) in Fig.~\ref{fig:shear_ens} show representative examples of linear-response trajectories and nonlinear trajectories exhibiting pronounced hysteresis and nonlinearity, respectively; the latter are excluded from the shear-modulus estimate. Raw data are shown as semi-transparent circles, while opaque squares represent averages over 100 consecutive values of $\gamma$ and $\sigma_{xy}^{\mathrm{tot}}$. Colors indicate the shear direction: forward shear $0 \rightarrow \gamma_m$ (red), reverse shear $\gamma_m \rightarrow -\gamma_m$ (green), and return $-\gamma_m \rightarrow 0$ (blue).

Fig.~\ref{fig:shear_ens}(d) shows the fraction of linear ensembles, $n_{\mathrm{lin}}/n_{\mathrm{tot}}$, as a function of the activity parameter $\Lambda$. It}
decreases with $\Lambda$, reaching zero at the melting point, $\Lambda = 10.0$ (Fig.~\ref{fig:shear_ens}(d)). This decline for $\Lambda > 6.0$ indicates activity-driven softening, marked by a departure from linear elasticity. Near $\Lambda = 10.0$, both $n_{\mathrm{lin}}$ and the shear modulus approach zero, signaling a fluidization transition with the complete loss of linear response.



%
\end{document}